# Capturing Lifecycle System Degradation in Digital Twin Model Updating

Yifan Tang[*]  
Email: yta88@sfu.ca

Mostafa Rahmani Dehaghani  
Email: mra91@sfu.ca

G. Gary Wang  
Email: gary_wang@sfu.ca

Product Design and Optimization Laboratory, School of Mechatronic Systems Engineering, Simon Fraser University, Surrey, BC, Canada

**ABSTRACT**

Digital twin (DT) has emerged as a powerful tool to facilitate monitoring, control, and other decision-making tasks in real-world engineering systems. Online update methods have been proposed to update DT models. Considering the degradation behavior in the system lifecycle, these methods fail to enable DT models to predict the system responses affected by the system degradation over time. To alleviate this problem, degradation models of measurable parameters have been integrated into DT construction. However, identifying the degradation parameters relies on prior knowledge of the system and expensive experiments. To mitigate those limitations, this paper proposes a lifelong update method for DT models to capture the effects of system degradation on system responses without any prior knowledge and expensive offline experiments on the system. The core idea in the work is to represent the system degradation during the lifecycle as the dynamic changes of DT configurations (i.e., model parameters with a fixed model structure) at all degradation stages. During the lifelong update process, an Autoencoder is adopted to reconstruct the model parameters of all hidden layers simultaneously, so that the latent features taking into account the dependencies among hidden layers are obtained for each degradation stage. The dynamic behavior of latent features among successive degradation stages is then captured by a long short-term memory model, which enables prediction of the latent feature at any unseen stage. Based on the predicted latent features, the model configuration at future degradation stage is reconstructed to determine the new DT model, which predicts the system responses affected by the degradation at the same stage. The proposed lifelong update method is evaluated with two real-world engineering datasets, e.g., battery degradation dataset from Oxford University, and flight engine degradation dataset from NASA. The test results demonstrate that the proposed update method could capture effects of system degradation on system responses during the lifecycle and outperform the conventional fine-tuning method in terms of prediction accuracy and robustness at future unseen stages.

## 1 Introduction

### 1.1 Literature review

Digital twin (DT) has been considered as a fundamental technology for health monitoring and process control during products' lifecycle, due to its ability to interact with physical entities (e.g., product, system). Generally, DT is defined as a virtual representation of a physical system [1, 2] that can mimic the time-series response of the physical system under various conditions and is capable of autonomous self-updating based on data acquired from the physical systems by sensing devices. Since the terminology was proposed in 2003, DT applications have been observed in manufacturing [3], process industry [4], aerospace [5], and other engineering tasks [6] to support collaboration, information access, and decision-making.

Before applying DT, historical data (e.g., sensor, simulation) of the physical system is used to construct an offline DT model via selected modeling methods, such as partial differential equations (PDEs) and data-driven models (e.g., artificial neural networks, support vector machines). Details on DT construction can be found in several literature reviews [2]. In most applications, historical data is collected from the early stage of its whole lifecycle, during which a practical system continues to change, such as operation conditions, accuracy of sub-systems, and functionality of system components. Consequently, the constructed DT model cannot represent the system accurately at each stage of the system's entire lifecycle. Therefore, after deploying the constructed DT model to the corresponding task, it is necessary to update the DT model with the data stream collected at each stage of the lifecycle, so that the DT model would function well in the whole lifecycle.

---

[*] Corresponding author.

According to the literature review on iterative updates of DT models [7], DT update methods could be classified according to the model types of DT. For PDEs-based DT models, the DT update methods are designed to tailor system parameters (e.g., mass or stiffness matrices) that are learned offline. This task is regarded as a parameter estimation problem, which has been solved by Kalman filters [8], Particle filters [9], and Bayesian estimation [10] and their various derivatives in data assimilation [11] and hybrid simulation [12]. For data-driven models, the DT update methods aim to tune model parameters by optimization, Bayesian estimation, or incremental computation [13], or modify the model structure using Bayesian techniques [14] or heuristic strategies [15]. Relevant works could be found in online learning [16], incremental learning [17], and lifelong learning [18]. Although DT update methods for both models have been studied, most works only make the updated DT model match the data collected at the current stage. In other words, the updated DT model would forget the system behavior at former stages in the lifecycle and will fail to forecast the system behavior accurately when the physical system changes at future stages.

One important fact is that a physical system degrades continuously in its lifecycle. Such system degradation could be the decreasing battery discharge capacity over charge-discharge cycles [19, 20], reduction of the remaining useful life in machines [21, 22] and other system health indicators that change over the lifecycle. To monitor a system's health, degradation models have been studied [23].

The first type of degradation model is physics-based. For instance, the Paris law was used to predict the crack growth of an operational unit [24, 25]. The Fick's diffusion law was adopted to model the corrosion rate in reinforced concrete [26]. The tunnel serviceability index was proposed to measure the service state of a tunnel infrastructure [27, 28] However, those physics models require prior knowledge of system physics and massive expensive experimental data for modeling. The applied assumptions restrict their generalizability in different applications.

With the increase in sensor measurements, data-driven models have emerged as the second type of degradation model. For instance, the physics-informed long short-term memory (LSTM) model has been used to predict the degraded battery capacities, which is applied to forecast the battery's remaining useful life [20]. The degraded stack voltage of a fuel cell could be obtained by a hybrid convolutional neural network-LSTM model from the outlet temperature and the historical stack voltage [29]. The dynamic health state of a bearing could be estimated from the sensor signals by an adaptive hidden Markov model [30]. The bi-directional gated recurrent unit model has been utilized to forecast the state of health for lithium-ion batteries based on charge temperature, voltage, and other observable properties [31]. The LSTM with Autoencoder is utilized to predict the remaining useful life of gas turbines directly from the operating and measurement signals [32]. A parallel hybrid model of finite element and deep neural network was proposed for bearings, where the deep neural network was trained to predict the maximum wear depth from vibration signals, and the finite element model was built to estimate the wear profile of bearings based on material, geometry, and load properties [33]. Both predictions are then used online to estimate the remaining useful life. Generally, those data-driven models are health prognostic models using the system responses as input.

The above works directly model system degradation as it happens. DT has also been used to predict system degradation. Current literature on DT applications in system degradation can be divided into two main categories, as shown in Figure 1. Explanations of the two categories of applications are given below.

*Category 1: Predict health indicators based on DT predictions of system responses.*

The most common application uses the DT model to predict time-series system responses, which are inputted into degradation models to forecast the remaining useful life and other health indicators [34]. For example, based on the incomplete discharge curve and the battery state-of-charge curve at one charge-discharge cycle, a LSTM model was applied to predict the completed discharge curve, and subsequently, the degraded capacity [35]. Similarly, the backpropagation neural work was adopted to predict the whole discharge curve [19]. The battery degradation state was then estimated by a convolutional neural network-LSTM-attention model from the predicted discharge curve and the historical discharge curves of previous cycles. For the proton exchange membrane fuel cell stack, the transformer neural network was adopted to predict the future stack voltage based on time-series operation parameters (e.g., temperature, current) and the observed voltage [36]. The remaining useful life is then estimated as the time when the future voltage is smaller than a threshold.

When obtaining data at one degradation stage, those models could be fine-tuned to provide accurate prediction at the current stage. However, system responses/signals at future degradation stages are unknown for practical applications, making them unable to forecast the degraded system responses for future stages in advance.

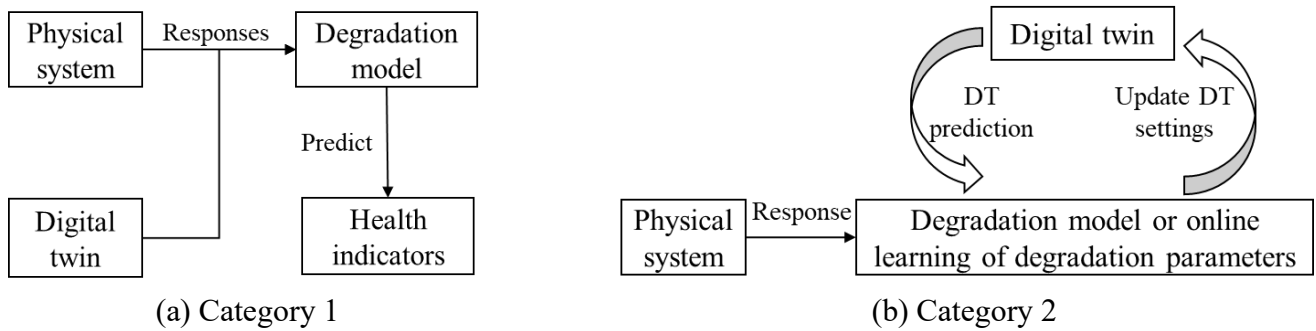

Figure 1 Categories of reviewed DT applications in system degradation.

*Category 2: Construct DT models with identified degradation parameters of physical systems.*

Instead of updating the entire DT model, some works construct DT models with identified degradation parameters, which are learned at each degradation stage. For instance, Simulink-based DT models were proposed based on identified degradation characteristics parameters (e.g., capacitance, inductance, parasitic resistance) for DC/DC power converters [37] and the DC/AC inverter [38]. At each degradation stage, those parameters are updated with actual data via Bayesian optimization or particle swarm optimization. Although both works can make the updated Simulink-based DT model precisely capture the actual system behavior at each single degradation stage, they cannot forecast the degraded system responses at future stages in advance, as the identified parameters continue degrading over the lifecycle.

To avoid the above limitation, degradation models for degradation parameters are trained based on historical data. The model can capture the trend of degradation parameters over the lifecycle, making the DT prediction more practical for future stages. For instance, a high-fidelity DT model was designed with finite element models for the electronically-controlled pneumatic brake system, whose degradation parameter is the direct-current resistance [39]. Based on the operating historical data, a model was trained to capture the degradation behavior of the resistance value. At any future stage, the degradation model is updated online to provide an accurate estimation of the direct-current resistance, which is used to update the initial parameter setting in the high-fidelity DT model. Similarly, a high-fidelity dynamic DT model with two degradation parameters, the pitting density and the wear depth, was proposed for the gearbox [40]. Based on historical data, a fatigue pitting model and an Archard wear model were trained to capture the propagation of the two degradation parameters [41]. During application, the comparison between responses from both the DT model and physical system was made to update the two degradation models, which would provide more practical degraded parameters. The obtained parameters were then utilized to update the initial geometry setting in the dynamic DT model for better prediction.

Apart from the above works which update degradation models during applications, some works pretrain an offline degradation model and apply the model without any further update. For instance, to predict the real-time temperature of lithium-ion batteries, a lumped thermal equivalent circuit (i.e., DT model) was designed with the heat capacity and the thermal generation rate [42]. The degradation of both parameters was captured by a LSTM model from the operation history data so that the DT model could update its model setting at each degradation stage for better prediction. Similarly, considering the thickness of the oxidation layer grows in semiconductor devices, a finite element model was designed based on affected mechanical properties, e.g., geometry and material behavior [43]. The experiments were conducted to model those properties as functions of aging, which makes the finite element analysis successful at each degradation stage.

Although those DT models could predict the degraded system responses in future stages in advance, they all require expert knowledge of the physical system to identify the degradation parameters before constructing the DT model. For a complex system, the degradation parameter would be hard to identify. Meanwhile, expensive experiments are necessary to collect sufficient data to construct the degradation model for identified parameters.

## 1.2 Motivation

Given a data-driven DT model and its corresponding physical system and assuming no domain expert knowledge, this paper proposes a lifelong DT update method to make the DT model work well at unseen future degradation stages. In other words, when the DT model is updated with online data from one degradation stage, the updated DT model has the capability to "memorize" the degradation effects on the system responses at former stages and predict system responses that would be affected by future degradation. The differences between the proposed method and the works in the current literature review are clarified below. Compared with current online DT update methods, the proposed method supports the updated DT model to predict the degraded time-series system responses at future stages in advance, with no information of future degradation. Compared with reviewed DT applications in system degradation, the proposed method would be inexpensive as it does not require the identified degradation parameters, prior knowledge of the physical system, or any expensive offline experiments, which would improve its applicability in different applications.

This work aims to address the above needs. Contributions of this paper are listed below.

- A novel DT update task is defined for systems with degradation during their lifecycle. Without any prior knowledge of the system and expensive experiments, the effects of system degradation on system responses are identified during the update process. According to the authors' knowledge, this specific DT update task is rarely or never discussed.
- A lifelong DT update method is proposed for DT constructed as a feedforward neural network. Different from conventional methods that update DT models stage by stage, the proposed method enables the prediction of DT configurations that are affected by system degradation at future unseen stages in the lifecycle.
- The proposed method provides an economical alternative to expensive aging tests, while also providing predictions of future unseen stages. Those predictions could be applied to system health management, which is not supported by predictions from the DT model using conventional fine-tuning method.

The remainder of the paper is structured as follows. The representation of system degradation in the DT model is discussed in Section 2.1, based on which the lifelong DT update method is discussed in detail in Section 2.2 and Section 2.3. The proposed method is then tested with the battery degradation dataset in Section 3 and the flight engine dataset in Section 4. Discussions are presented in Section 5, after which Section 6 summarizes the work in the paper.

## 2 Methods

## 2.1 Representation of system degradation in DT

Degradation in physical systems reflects a continuous and gradual system performance deterioration over the lifecycle [44]. Such continuous performance change indicates that even under the same operation setting, time-series responses of the same physical system differ among degradation stages in the whole lifecycle. Capturing these dynamics necessitates an approach that allows for the seamless representation of degradation effects on the system responses in the lifecycle.

In this paper, the prior knowledge and degradation parameters of physical systems are assumed unknown. DT models are constructed only based on simulation data or experimental data at the early stage of the lifecycle via machine learning models, such as neural networks. Figure 2 depicts the conventional fine-tuning framework to update the data-driven DT model through the lifecycle of most applications. If a DT model with configuration $\boldsymbol{\theta}_0$ is obtained based on data collected from the physical system at the initial degradation stage $T_0$ (e.g., no system degradation), the DT model would be updated based on data from the next degradation stage $T_1$. The updated DT model with the new configuration $\boldsymbol{\theta}_1$ could then simulate the degraded system at the stage $T_1$. With the system degrading continuously in the lifecycle, such an update process is repeated at each degradation stage. Therefore, the obtained DT configuration $\boldsymbol{\theta}_i$ at the degradation stage $T_i$ could be regarded as a representation of the effect of system degradation on system responses at the stage $T_i$. Compared with reviewed works relying on degradation parameters, the selected degradation representation only depends on DT model configuration, making the defined representation more general and applicable to different physical systems.

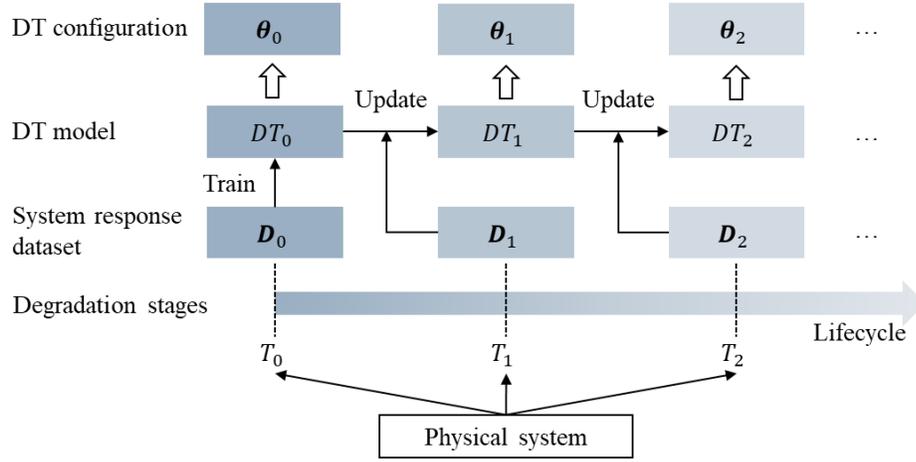

Figure 2 Conventional fine-tuning framework for a DT model.

## 2.2 Lifelong DT update method

As mentioned in Section 1.2, this paper aims to propose a lifelong DT update method, so that the update DT model is able to predict the time-series system future responses directly affected by system degradation. To fulfill this purpose, a model that captures the dynamic system degradation over the lifecycle is required, such as the reviewed works using the degradation parameters or degradation model [37–43]. Instead of using prior knowledge, constructing a dynamic model of DT configurations over degradation stages would be a promising counterpart, according to the defined degradation representation. More specifically, if a dynamic model $\boldsymbol{\theta}_i = f(\cdot, T_i)$ is learned during the lifelong update process, DT configurations at future degradation stages could be predicted directly, thereby capturing the system responses affected by degradation at future stages.

By integrating the tuning process and the dynamic model construction, the lifelong DT model update method is proposed as shown in Figure 3. The detailed steps are described below.

***Step 1 (Initialization phase)***: Given a physical system, its initial data-driven DT model $DT_0$ is constructed on a dataset $\boldsymbol{D}_0$ obtained at the stage $T_0$, where no degradation is assumed. The obtained DT configuration $\boldsymbol{\theta}_0$ is stored in the configuration database, based on which the dynamic model structure is determined and fixed during the future lifelong update process.

***Step 2 (Warm-up phase):*** In most real-life scenarios, the physical system would work well during the early stage of the lifecycle. In this paper, a group of $m$ earlier degradation stages $\{T_1, …, T_m\}$ is regarded as the warm-up phase, where the DT model is continuously fine-tuned based on the system response dataset collected at each degradation stage as shown in Figure 2. Configurations of the updated DT model at each stage are saved in the configuration database. Based on the stored configurations $\{\boldsymbol{\theta}_0, \boldsymbol{\theta}_1, \ldots, \boldsymbol{\theta}_m\}$, the dynamic model $\boldsymbol{\theta}_i = f(\cdot, T_i)$ is trained with predefined structures, whose details are presented in Section 2.3.

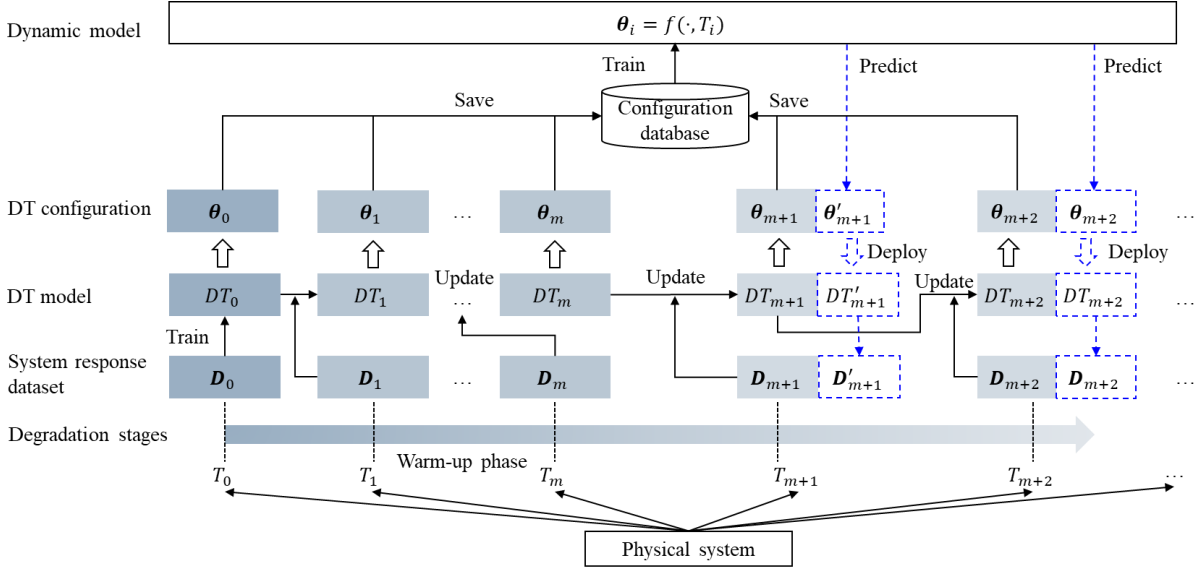

Figure 3 Proposed lifelong DT update method. The solid lines and arrows indicate the lifelong update process, and the dashed lines and arrows refer to the application of the DT.

**Step 3 (Lifelong application phase)**: For any future degradation stage $T_i$ ($i > m$), the trained dynamic model $f$ could be adopted to predict the DT configuration $\boldsymbol{\theta}'_i$, based on which the estimated DT model $DT'_i$ is obtained. The system responses affected by degradation at future stages $\{T_i, \ldots, T_{i+j}\}$ are then predicted by corresponding estimated DT models $\{DT'_i, \ldots, DT'_{i+j}\}$. The estimated system responses $\{\boldsymbol{D}'_i, \ldots, \boldsymbol{D}'_{i+j}\}$ could be applied to calculate the health state of the system for management and other services, which will not be covered in the paper.

**Step 4 (Lifelong update phase)**: After completing the task at the $i$-th degradation stage $T_i$, the actual dataset $\boldsymbol{D}_i$ is obtained to update the DT model $DT_{i-1}$ with configuration $\boldsymbol{\theta}_{i-1}$. The updated model is denoted as $DT_i$, whose configuration is $\boldsymbol{\theta}_i$. Compared with the estimated configuration $\boldsymbol{\theta}_i'$ from the previous dynamic model, the new configuration $\boldsymbol{\theta}_i$ captures the actual effect of degradation on system responses at the stage $T_i$. To improve the accuracy of the proposed dynamic model gradually, the new configuration $\boldsymbol{\theta}_i$ is stored in the configuration database to retrain the dynamic model. More details are discussed in Section 2.3.

## 2.3 Dynamic model for DT configuration

Generally, the configuration of a data-driven DT model includes the structure and parameters, which could be updated simultaneously, such as the works in [14, 15] In this paper, a feedforward neural network (FNN) with a fixed structure is adopted as the DT model, meaning that the dynamic model only needs to capture how the DT model parameters change during the whole lifecycle.

The general framework of the proposed dynamic model is summarized in Figure 4, where Autoencoder is adopted to capture the latent features of DT configurations at each degradation stage, and LSTM is used for time-series forecasting of future configuration features based on features at previous $w$ degradation stages. In the following discussion, the proposed dynamic model is denoted as Autoencoder-LSTM for simplification.

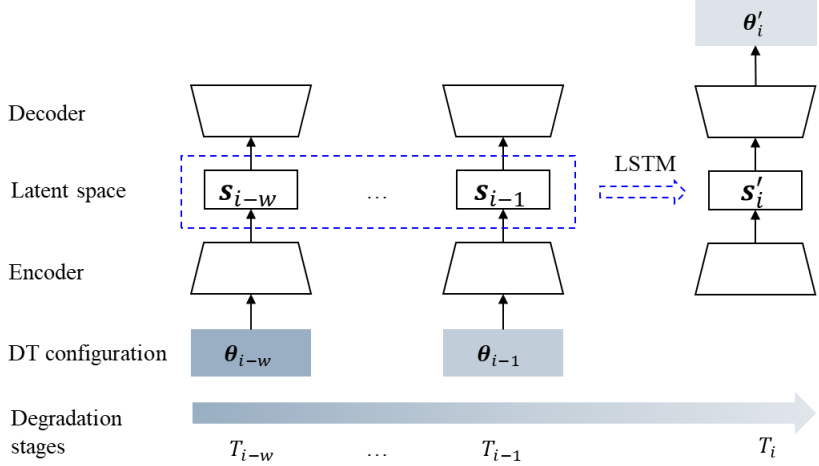

Figure 4 Framework of the proposed dynamic model of DT configuration.

### 2.3.1 Autoencoder to reconstruct model configurations

Considering a FNN model with $N$ hidden layers $\{l_1, \dots, l_N\}$, the input dimension and the output dimension of the $i$-th hidden layer $l_i$ are $n_{in}^i$ and $n_{out}^i$ respectively. In this work, each hidden layer uses a linear transformation, e.g., $y = xA^T + b$, where $A$ is the weight matrix and $b$ is the bias vector. Therefore, the number of parameters at the hidden layer $l_i$ is $(n_{in}^i + 1) \times n_{out}^i$. The total number of parameters in a model configuration $\boldsymbol{\theta}$ is $\Sigma_{i=1}^{N}(n_{in}^i + 1) \times n_{out}^i$. This number could be thousands, millions, or even more, making the task of capturing the dynamic behavior of model configuration a high-dimensional problem. If each parameter is learned by a separate model, a large memory is required to store thousands of models. If a simple FNN is used to predict the configuration based on the degradation stage, the output dimension is then huge, making the training process harder to converge. Both options are thus not ideal and may risk poor prediction performance at future degradation stages.

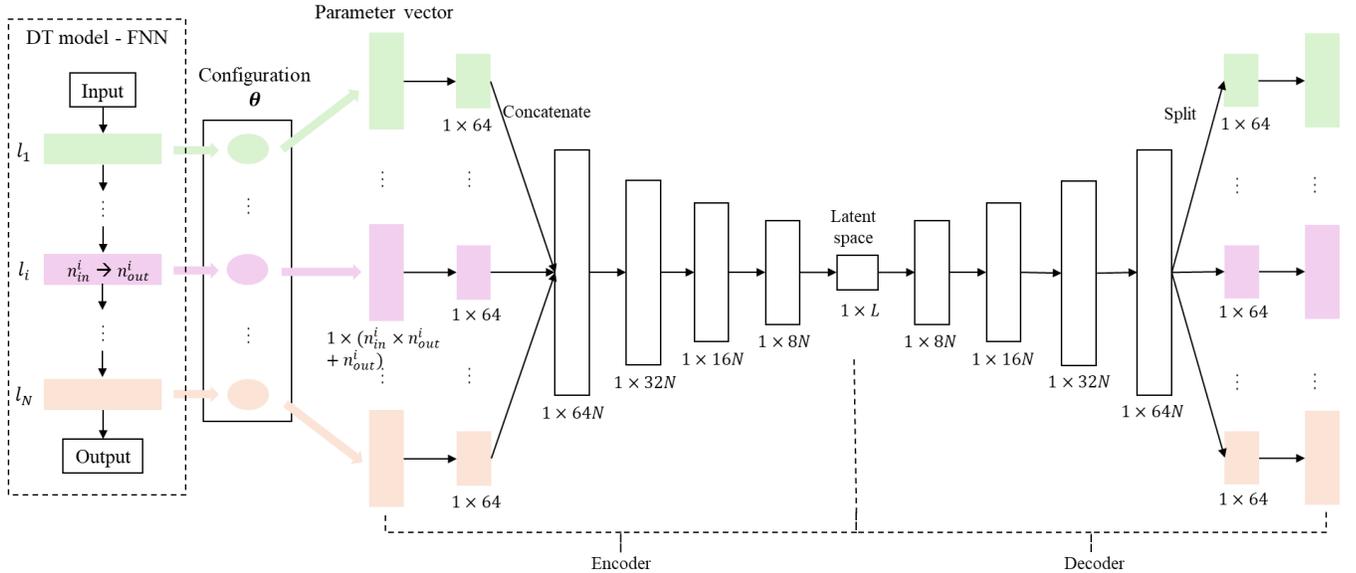

Figure 5 Autoencoder structure to reconstruct the high-dimensional configurations.

To reduce the problem dimension in this paper, Autoencoder is utilized to represent the model configuration $\boldsymbol{\theta}$. The encoder learns efficient low-dimensional features, based on which the decoder reconstructs the entire model configuration. Autoencoder has been a popular tool for feature extraction in deep learning. Compared to traditional dimension reduction techniques (e.g., proper orthogonal decomposition), Autoencoder can capture complex nonlinear relationships. Different from supervised learning approaches, Autoencoder does not require labeled data, which is advantageous when dealing with large-scale model configurations where annotations are unavailable. Based on those merits, Autoencoder has been applied in fluid dynamics [45], material [46], and other engineering fields [47].

Figure 5 shows the proposed Autoencoder structure with details summarized in Table 1. Given the model configuration $\boldsymbol{\theta}$, the weights and biases of each hidden layer $l_i$ are flattened as a parameter vector with the size $1 \times \left(\left(n_{in}^i + 1\right) \times n_{out}^i\right)$. The encoder takes parameter vectors of all hidden layers as inputs simultaneously, aiming to capture the inherent relationship among all hidden layers. To alleviate the effects of various dimensions, each parameter vector is compressed to a vector with the same size $1 \times 64$ by the corresponding linear hidden layer. The compressed vectors of $N$ hidden layers are then concatenated to one vector with the size $1 \times 64N$, which are processed by four hidden layers with output dimensions $1 \times 32N$, $1 \times 16N$, $1 \times 8N$, and $1 \times L$ respectively. In the latent space, $L$ is the dimension of configuration features extracted from the input model configuration. Once the configuration features are obtained, they are fed to the decoder, where four hidden layers are adopted to output the concatenated vector with the size $1 \times 64N$. The vector is then divided to $N$ separate vectors with the same dimension $1 \times 64$. Each split vector is processed by a linear hidden layer to reconstruct the parameter vector in the DT model. The reconstructed parameter vectors are then deployed to the DT model for system response prediction.

Table 1 Details of the proposed Autoencoder model

| # | Layers | Activation | Input dimension | Output dimension |
|---|---|---|---|---|
| 1 | $N \times$Linear() | Tanh($\cdot$) | $1 \times \left(\left(n_{in}^i + 1\right) \times n_{out}^i\right), i \in [1, N]$ | $N \times 1 \times 64$ |
| 2 | Concatenation | | $N \times 1 \times 64$ | $1 \times 64N$ |
| 3 | Linear() | Tanh($\cdot$) | $1 \times 64N$ | $1 \times 32N$ |
| 4 | Linear() | Tanh($\cdot$) | $1 \times 32N$ | $1 \times 16N$ |
| 5 | Linear() | Tanh($\cdot$) | $1 \times 16N$ | $1 \times 8N$ |
| 6 | Linear() | | $1 \times 8N$ | $1 \times L$ |
| 7 | Linear() | Tanh($\cdot$) | $1 \times L$ | $1 \times 8N$ |
| 8 | Linear() | Tanh($\cdot$) | $1 \times 8N$ | $1 \times 16N$ |
| 9 | Linear() | Tanh($\cdot$) | $1 \times 16N$ | $1 \times 32N$ |
| 10 | Linear() | Tanh($\cdot$) | $1 \times 32N$ | $1 \times 64N$ |
| 11 | Split | | $1 \times 64N$ | $N \times 1 \times 64$ |
| 12 | $N \times$Linear() | | $N \times 1 \times 64$ | $1 \times \left(\left(n_{in}^i + 1\right) \times n_{out}^i\right), i \in [1, N]$ |

When designing Autoencoder, selecting an appropriate dimension $L$ of the latent features is fundamental, as it affects the model performance in terms of data reconstruction, feature extraction, and generalization [48]. If the dimension is too low, Autoencoder may not capture essential features, resulting in loss of information. Conversely, an excessively large latent space would cause overfitting and fail to capture the underlying data structure. The most applied method involves training Autoencoder with varying latent dimensions and evaluating their impact on downstream tasks (e.g., classification accuracy) and reconstruction error, which is computationally intensive [49]. The dimension of latent features could also be determined by the singular value decomposition method where dimensions of singular values above a threshold are retained [50], or by incorporating the regularization term of latent space during the training process to encourage the encoder utilizing fewer latent features [51]. However, the performance of both methods depends on the dataset, model structure, and downstream task.

Instead of finding the optimal dimension, this paper determines the dimension $L$ of latent features based on the information theory and the model configuration $\boldsymbol{\theta}_0$ at the initial degradation stage $T_0$. The dimension is then fixed during the lifelong update process. In data compression tasks, the Shannon entropy is regarded as the minimum number of bits to compress a group of random variables [52, 53] According to [54], the stochastic behavior of model parameters during the training process could be considered as a Gibbs distribution. Assuming that each model parameter could be regarded as sampled randomly from the same distribution, the probability of each model parameter could be estimated as:

$$p_{l_i}^j = \frac{\exp(-\beta \theta_{0,l_i}^j)}{\sum_i^N \sum_j^{(n_{in}^i+1) \times n_{out}^i} \exp(-\beta \theta_{0,l_i}^j)} \tag{1}$$

where $\theta_{0,l_i}^j$ is the $j$-th model parameter of the $i$-th hidden layer $l_i$ at the degradation stage $T_0$, $p_{l_i}^j$ is the corresponding probability, and $\beta$ is a constant value. The Shannon entropy of model configuration $\boldsymbol{\theta}_0 = \{\theta_{0,l_i}^j | i \in [1, N], j \in [1, (n_{in}^i + 1) \times n_{out}^i]\}$ is then calculated as Eq. (2).

$$H = -\sum_{i,j} p_{l_i}^j \log_2 p_{l_i}^j \tag{2}$$

The dimension $L$ is finally determined as:

$$L = a \times [H] \tag{3}$$

where $[H]$ is the integer number closest to the entropy $H$. Considering that the model configuration evolves continuously during the system lifecycle, the entropy of model configurations varies at different degradation stages. Therefore, $a > 1$ is a ratio to enlarge the integer value, which would reduce the risk of the low-dimensional latent space failing to capture the underlying features at future degradation stages. In the paper, the hyperparameters are set as $\beta = 1$ and $a = 2$ when testing the two engineering datasets.

### 2.3.2 LSTM to predict configuration features

After training the autoencoder, the learned latent features $s_i$ of model configurations at all known degradation stages are accessible. When using the learned autoencoder to predict configurations, the autoencoder structure is fixed and the prediction performance only depends on the estimated latent features. As discussed in Section 2.1, the system degradation is represented as the dynamic behavior of latent features over the lifecycle. Therefore, a time-series forecasting model is required to estimate the latent features at future degradation stages.

In the time-series forecasting tasks, LSTM has been a powerful tool due to its ability to capture long-term dependencies in sequential data while mitigating the vanishing gradient problem [55]. Unlike traditional recurrent neural networks, LSTM incorporates gating mechanisms, i.e., input, forget, and output gates, which selectively retain relevant information and discard irrelevant information. This characteristic makes LSTM effective in capturing complex temporal patterns over sequential data, such as financial market trends, weather prediction, and industrial processes. Furthermore, LSTM outperforms classical time-series models by handling non-linearity and non-stationary time-series data without requiring extensive feature engineering. Based on these advantages, LSTM has been observed in a wide range of applications, including financial market forecasting [56], , healthcare analytics [57], and anomaly detection [58].

Based on the above discussion, this paper adopts the LSTM as the time-series forecasting model for latent features of model configurations. In theory, an optimal time-series forecasting model exists for the proposed task, where the number of hidden layers, dimension of hidden states in the LSTM cell, activation function, and other hyperparameters are optimized by hyperparameter optimization [59]. As hyperparameter optimization is not the research topic of this work, the proposed model structure is determined by trial and error, instead of the expensive optimization.

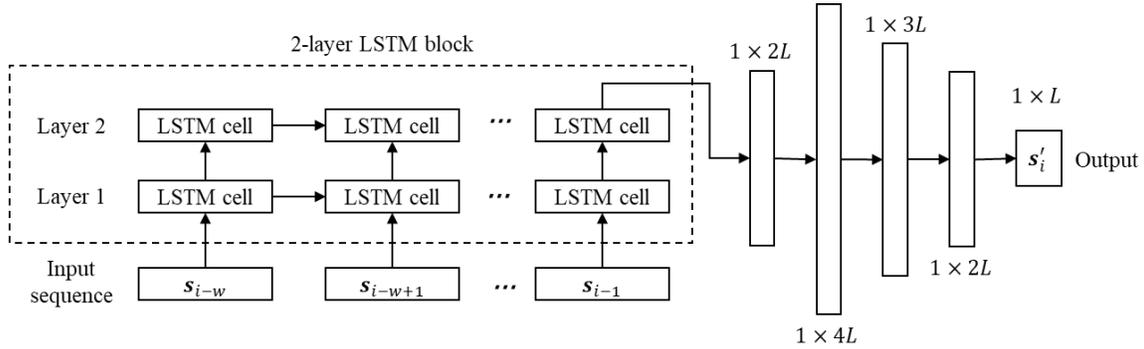

Figure 6 LSTM structure to predict the configuration feature.

Table 2 Details of the proposed LSTM model.

| # | Layers | Activation | Input dimension | Output dimension |
|---|---|---|---|---|
| 1 | LSTM block | | Sequence $w \times (1 \times L)$ | $1 \times 2L$ |
| 2 | Linear() | Tanh(·) | $1 \times 2L$ | $1 \times 2L$ |
| 3 | Linear() | Tanh(·) | $1 \times 2L$ | $1 \times 4L$ |
| 4 | Linear() | Tanh(·) | $1 \times 4L$ | $1 \times 3L$ |
| 5 | Linear() | Tanh(·) | $1 \times 3L$ | $1 \times 2L$ |
| 6 | Linear() | | $1 \times 2L$ | $1 \times L$ |

Figure 6 and Table 2 summarize the model structure and details, respectively. In the model, a two-layer LSTM block is adopted. The input sequence $s_{i-w}, \ldots, s_{i-1}$ with each element having the dimension of $1 \times L$ is processed by the first LSTM layer, whose output hidden states are input to the second LSTM layer. The process continues till the last LSTM layer. In this work, each LSTM cell is implemented with the hidden state size $1 \times 2L$ using the Pytorch library and other hyperparameters are at their default values. The output state of the last cell in the second LSTM layer is then processed by several hidden layers with the activation function Tanh(·) to obtain the final output as the predicted latent feature vector $s'_i$ with the size $1 \times L$.

Once the proposed LSTM model has been trained, a progressive prediction framework is adopted to predict latent features of model configurations at all future unknown degradation stages. As shown in Figure 7, if we have known all latent features before the $i$-th degradation stage and want to predict all latent features at each future stage, an input matrix $[s_{i-w+1}; \ldots; s_i]$ with the window size $w$ is fed to the trained LSTM model to predict the latent feature $s'_{i+1}$ at the $(i+1)$-th degradation stage. Then the updated input matrix $[s_{i-w+2}; \ldots; s_i; s'_{i+1}]$ is adopted to obtain the estimated latent feature $s'_{i+2}$ at the $(i+2)$-th degradation stage. This process will repeat until the latent feature at each future degradation stage is obtained. Based on those predicted latent features, the estimated model configuration at each future degradation stage is obtained by the decoder in the trained autoencoder model, which would be applied to predict the system responses affected by the system degradation.

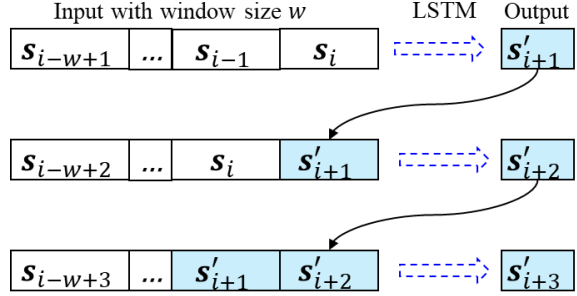

Figure 7 Prediction of future configuration features based on features in observed degradation stages.

## 3 Oxford Battery Degradation Dataset

### 3.1 Dataset description and preparation

The battery dataset is collected by the Battery Intelligence Lab at Oxford University [60]. The dataset contains measurements from the eight SLPB533459H4 batteries produced by Kokam CO LTD, which are tested by Bio-Logic MPG-205 with eight channels. During the test, the battery works under a constant-current-constant-voltage charging profile, followed by a drive cycle discharging profile. The charging-discharging rate is 1C, which means the battery is charged/discharged at a current equal to its rated capacity. The specific basic parameters of the battery are summarized in Table 3. Among the eight cells in the battery, only the first cell is studied in the test.

Table 3 Parameter of the tested battery in the Oxford battery degradation dataset (Revised from [19])

| Parameter | Value | Unit |
| --- | --- | --- |
| Constant current | 740 | mA |
| Rated capacity | 740 | mAh |
| Voltage range | 2.7-4.2 | V |
| Environmental temperature | 40 | °C |
| Cell number | 8 | |

In the dataset, the battery is charged and discharged repeatedly until the battery capacity reduces to a specified value when the battery meets the retirement requirement. As the service life of the battery is quite long, the battery properties (e.g., time, voltage, charge, and temperature) are measured and stored every 100 charging-discharging cycles. The degradation stages in Section 2.2 are then defined accordingly, e.g., regarding the cycle 0 as the stage $T_0$, the cycle 100 as the stage $T_1$, and the cycle 8200 as the stage $T_{82}$. Based on the stored measurements, Figure 8 summarizes degradation curves in Cell 1 of the battery over charging-discharging cycles (i.e., the lifecycle). According to Figure 8 (a), the battery capacity decreases obviously with the charging-discharging cycle number, indicating that the battery health degrades continuously during the lifecycle. Affected by the battery capacity degradation, the time to charge the battery from 2.7V to 4.2V and the time to discharge the battery from 4.2V to 2.7V reduce with the cycle number, as shown in Figure 8 (b) and Figure 8 (c) respectively.

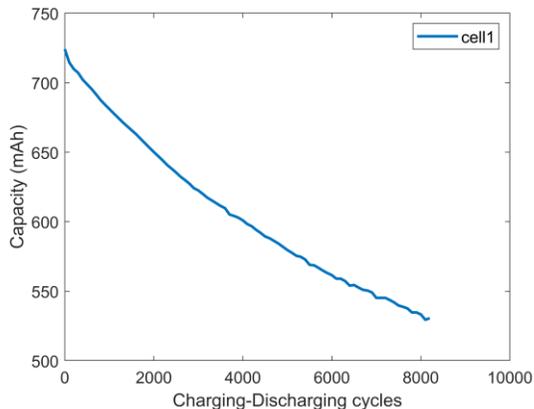
(a) Discharge capacity over cycles

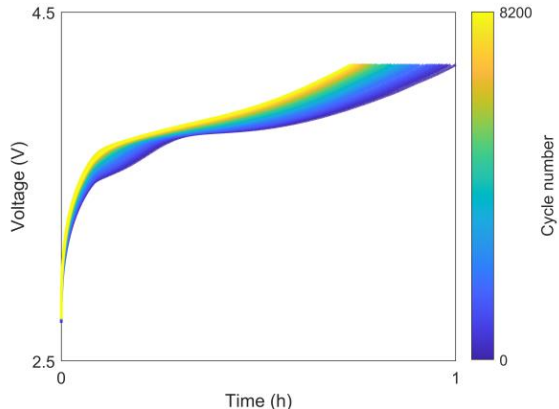
(b) Voltage curves during the charging

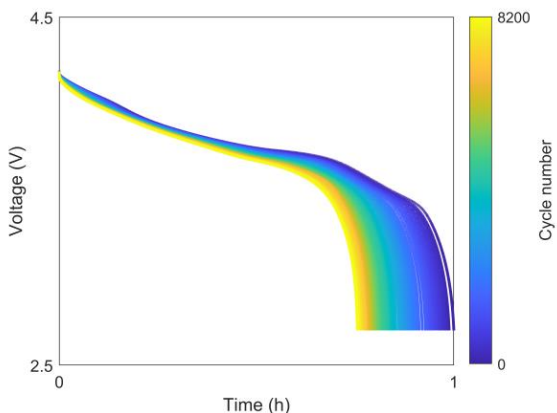
(c) Voltage curves during the discharging

Figure 8 Battery degradation curves.

In this test, the proposed lifelong update method aims to capture the effect of the degraded capacity on the voltage curves and enable the DT model to forecast the affected voltage curves at all future unseen charging-discharging cycles. More specifically, two FNN models are used as DT models for the charging voltage curve and the discharging voltage curve. According to the defined task in Section 1.2, the partial voltage curves at future charging-discharging cycles are unknown in advance in this work. This is different from works reconstructing the whole voltage curve based on a partial curve, which only works at the known stages [35, 36]. Therefore, the inputs of both FNN models are defined only based on the predefined charging or discharging settings, i.e., constant current $c = 0.74A$, and time $t$ ($s$). The output is the voltage $v$ at the time $t$ when charging or discharging with a constant current. Considering that the number of data points on each curve is thousands at each cycle in the dataset, 200 data points, i.e., voltages with corresponding time steps, are sampled evenly from the raw voltage curves to simply the modeling task. In other words, the charging and discharging curves at each cycle contain 200 data points respectively. Finally, datasets covering 80 cycles (200 data points on each cycle) are obtained for both charging and discharging curves, as the data from cycles 3400, 4700, and 4900 (i.e., degradation stages 34, 47, and 49) are not given in the published dataset.

### 3.2 Test setting

The proposed lifelong update process consists of several training components, including the training of initial DT models $DT_0$, the fine tuning of DT models based on data from each degradation stage, and the training of dynamic model for the model configuration during the lifecycle. Detailed settings for those parts are discussed below.

During this test, the DT models for the charging and discharging curves share an identical structure, as shown in Figure 9. As the nonlinearity in both charging and discharging curves is low, a simple structure using four hidden layers is utilized to predict the voltage from the constant current and time. All hidden layers use the same activation function Tanh(·). Based on 200 data for the charging cure and 200 data for the discharging curve at the initial degradation stage $T_0$, the model $DT_0$ is trained with the detailed hyperparameters in Table 4. The epoch number is 1,000. The loss function is the mean square error (MSE) between the predicted and actual outputs, which is minimized by the Adam optimizer with the learning rate 0.005. Figure 10 summarizes the training process of $DT_0$ based on 200 data points from the initial stage $T_0$, where 80%, 10%, and 10% of data are used for training, validation, and testing respectively. As their MSE values converge to around 0, the predefined FNN structure in Figure 9 could capture the charging and discharging behavior of the battery.

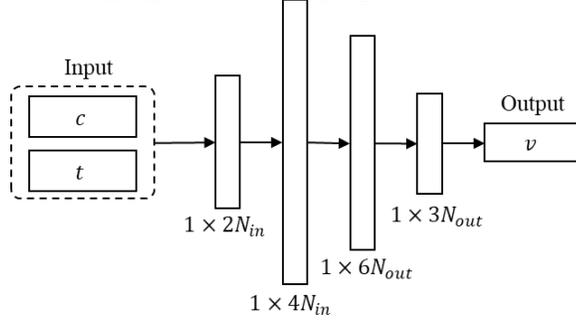

Figure 9 FNN structure for battery dataset. $N_{in} = 2$ and $N_{out} = 1$ are the input and output dimensions of the model respectively. The model input contains a constant current $c$ and time $t$. The output is the voltage $v$.

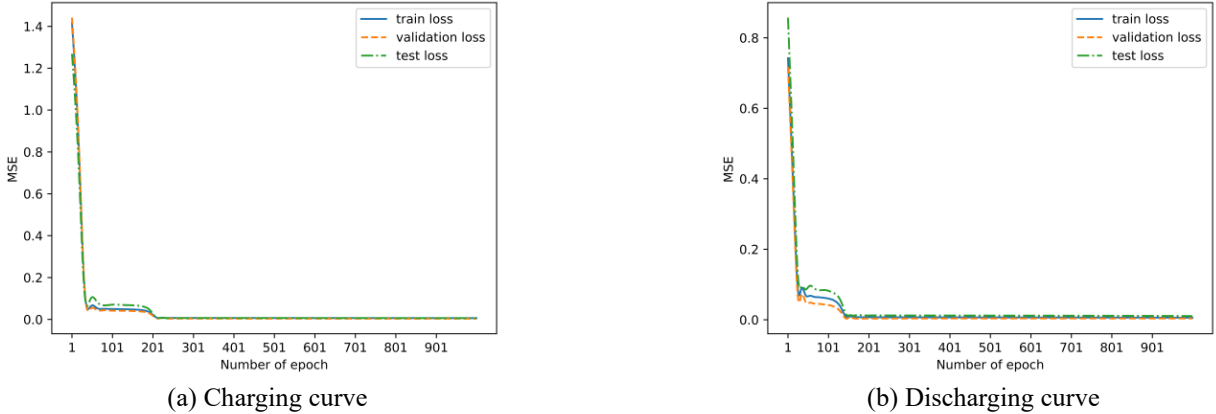

(a) Charging curve  (b) Discharging curve

Figure 10 Training process of $DT_0$ for battery dataset.

Once the initial DT model is trained, the DT model is fine-tuned sequentially in the warm-up phase, which consists of $m = 20$ degradation stages, i.e., $\{T_1, \ldots, T_{20}\}$. At each stage, all collected data is used to update the DT model, with the epoch number 10, the learning rate 0.001, and the optimizer Adam. Different from training the initial DT model, the loss function for the fine-tuning process is defined as below,

$$loss = MSE + \alpha \times l^2(\cdot) \quad (4)$$

where $l^2(\cdot)$ is the 2-norm regularization term on model parameters, so that the model parameters would not change drastically after tuning and the overfitting risk is reduced. $\alpha$ is the coefficient for the regularization term, which is set as $10^{-5}$ in the test.

In both charging and discharging, the entropy values of $\boldsymbol{\theta}_0$ are around 6.9 and the dimensions of latent space in Autoencoders are calculated as 14 during the test. When training the proposed autoencoder based on the stored model parameters, the epoch number is set to 1,000, and the learning rate is 0.0001. The loss function is the same as Eq. (4) to maintain the parameter stability in the deep structure, where the coefficient for regularization term is $10^{-4}$ in the test. The optimizer is still Adam. Once the latent features of model parameters are captured from the trained autoencoder, the LSTM is trained with the same hyperparameter setting as the autoencoder.

Table 4 Hyperparameter setting when testing the lifelong update method with the battery dataset.

|  | Hyperparameter | Value |
| --- | --- | --- |
|  | Number of stages in warm-up phase, $m$ | 20 |
|  | Ratio to enlarge entropy of model parameters, $a$ | 2 |
|  | Constant value in Gibbs distribution, $\beta$ | 1 |
|  | Window size for input sequence of LSTM, $w$ | 5 |
| Train $DT_0$ | Epoch number | $10^3$ |
|  | Learning rate | $5 \times 10^{-3}$ |
|  | Loss function | $MSE$ |
|  | Optimizer | Adam |
| Find tune DT models | Epoch number | 10 |
|  | Learning rate | $10^{-3}$ |
|  | Loss function | $MSE$, $l^2$-norm regularization |
|  | Weight of $l^2$-norm in the final loss, $\alpha$ | $10^{-5}$ |
|  | Optimizer | Adam |
| Train Autoencoder and LSTM | Epoch number | $10^3$ |
|  | Learning rate | $10^{-4}$ |
|  | Loss function | $MSE$, $l^2$-norm regularization |
|  | Weight of $l^2$-norm in the final loss, $\alpha$ | $10^{-4}$ |
|  | Optimizer | Adam |

According to the authors' best knowledge, the online update task proposed in this work, where the effect of system degradation is modelled in a DT update process, has not been discussed in current literature review. To demonstrate the effectiveness of the proposed lifelong update method, the conventional fine-tuning framework in Figure 2 is adopted for comparison. The hyperparameter setting for the conventional fine-tuning framework is the same as the one for fine-tuning DT models in the proposed lifelong update method.

### 3.3 Results

Based on the above processed datasets from 80 degradation stages and the defined DT model structure, the proposed lifelong update method is evaluated. According to the description in Section 2.2, when the actual data from the $i$-th degradation stage $T_i$ is obtained during the lifelong update phase, the previous DT model $DT_{i-1}$ is fine-tuned to get the new model $DT_i$. The new DT configuration $\boldsymbol{\theta}_i$ is stored with all previous configurations to retrain the Autoencoder-LSTM model, which is used to predict the DT model $DT_j'|j>i$ at any future $j$-th future degradation stage. In such cases, two types of predictions could be obtained for the $j$-th stage, i.e., the one obtained from the model $DT_i$, and the one obtained from the model $DT_j'$. Compared with the actual data at the $j$-th stage, the MSE value calculated from the former prediction indicates the performance of the fine-tuned DT model $DT_i$, while the MSE value calculated from the latter prediction reflects the performance of the degraded model $DT_j'$ generated from the proposed lifelong update method. Therefore, after completing the update process at the $i$-th stage, MSE values at all future degradation stages could be obtained for the fine-tuned DT model and the Autoencoder-LSTM generated DT. To clarify the following discussion, the Autoencoder-LSTM generated DT at the $i$-th stage is defined as a group of estimated DT models $\{DT_j'|j=i+1, i+2, ...\}$ after completing the proposed lifelong update process at the same stage.

During the implementation of the test, the last five degradation stages are not used for online update, so that the DT performance could still be evaluated when completing the last update step. Figure 11 summarizes MSE values at all future degradation stages obtained by fine-tuned and Autoencoder-LSTM generated DT models when completing the update at each degradation stage during the lifelong update phase. For example, after completing the first lifelong update at the $20^{th}$ stage, all MSE values at future stages obtained by Autoencoder-LSTM generated DT are shown as the dark blue curve starting from the $21^{st}$ stage in Figure 11 (b). The MSE values obtained after completing the last lifelong update are plotted as the lowest yellow curves. Compared with the MSE value curves obtained by the fine-tuned DT, the overall range of MSE values at each future degradation stage is smaller in the Autoencoder-LSTM generated DT in most cases, although some outliers are observed at the first several update steps. The potential reason may be the insufficient predefined warm-up phase, resulting in the proposed Autoencoder-LSTM failing to capture the effects of the system degradation efficiently.

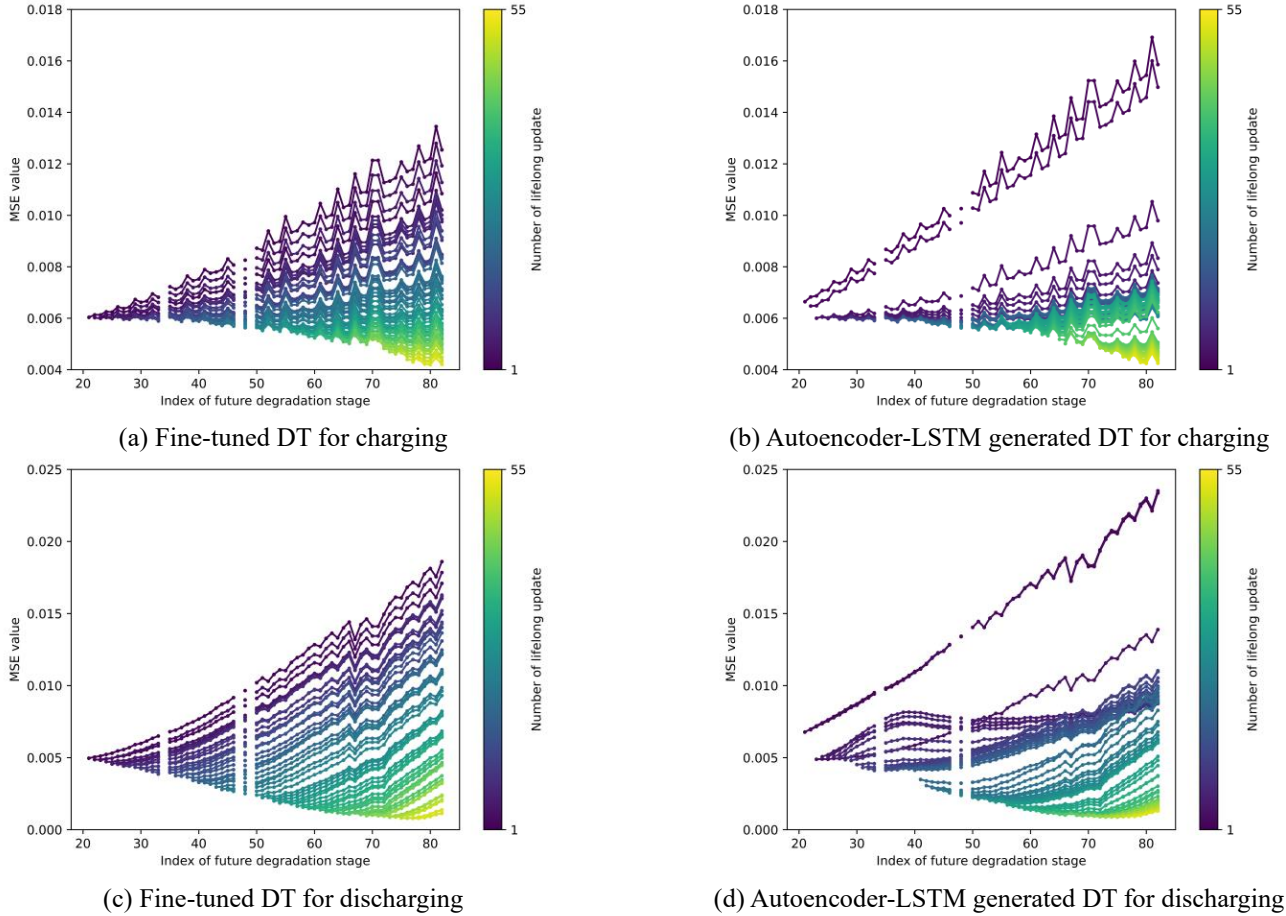

(a) Fine-tuned DT for charging  (b) Autoencoder-LSTM generated DT for charging

(c) Fine-tuned DT for discharging  (d) Autoencoder-LSTM generated DT for discharging

Figure 11 MSE values of the fine-tuned and Autoencoder-LSTM generated DT models with the battery dataset.

To clarify the performance difference at each degradation stage between fine-tuned and Autoencoder-LSTM generated DT models, Figure 12 summarizes the boxplot of MSE values obtained at each future stage during the whole lifelong update phase except for the first 10 update steps. More specifically, Figure 12 is another representation of the same performance dataset in Figure 11. For instance, the two MSE values at the $32^{nd}$ stage are collected from each DT after completing the $11^{th}$ lifelong update step at the $30^{th}$ stage and the $12^{th}$ lifelong update step at the $31^{st}$ stage. The number of MSE values at $i$-th stage in the figure could be calculated as $i - 10 - m$, where $m$ is the number of stages in the warm-up phase. The figures indicate that for both battery charging and discharging datasets, the Autoencoder-LSTM generated DT outperforms the fine-tuned DT models at most future degradation stages, in terms of smaller median MSE values (i.e., the orange line in the box) and smaller box sizes. The comparison results demonstrate that during the lifelong update phase, the proposed method can capture the effect of system degradation on the system responses at future unseen degradation stages in advance, better than the fine-tuned DT with smaller variations.

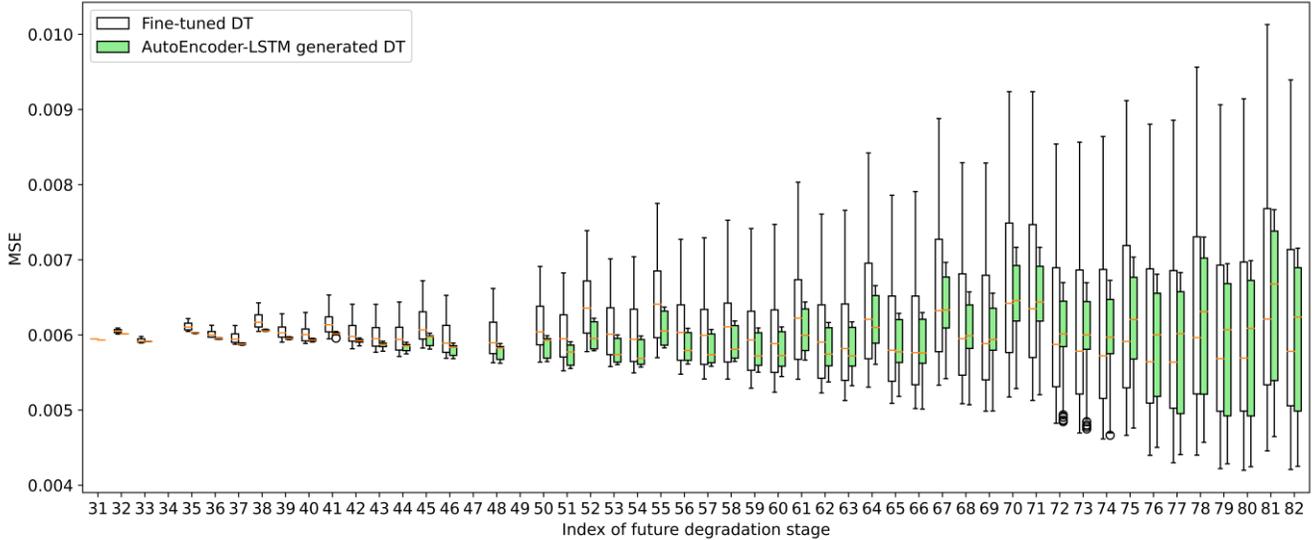

(a) Charging

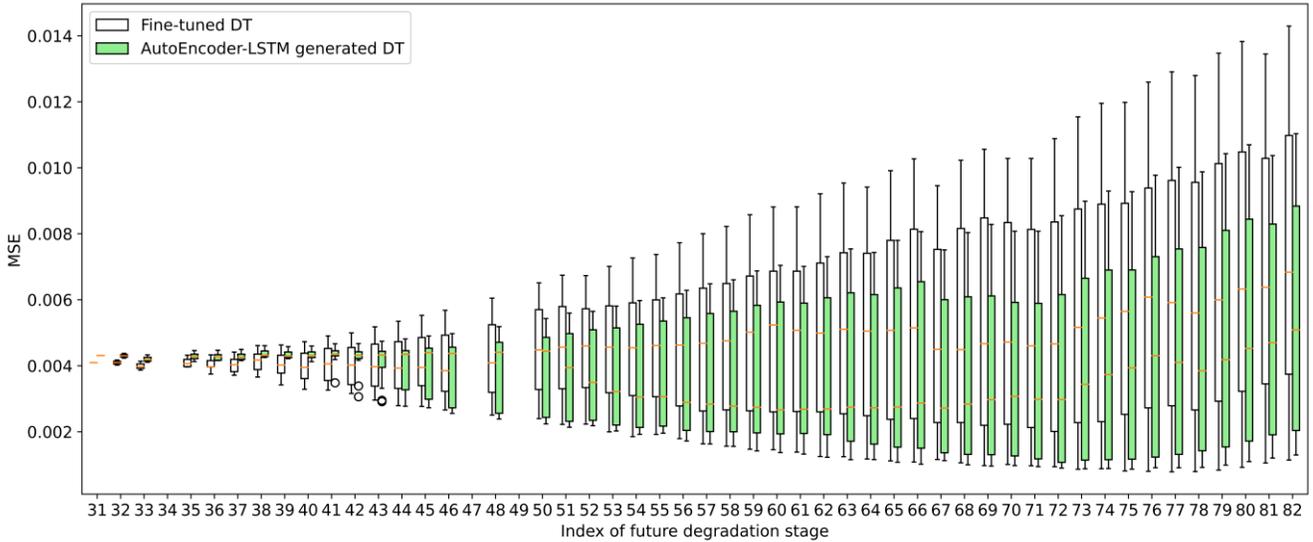

(b) Discharging

Figure 12 Boxplot of MSE values at each stage during the lifelong update phase with the battery dataset.

Apart from the above performance comparison, a success ratio is defined to compare the performance of fine-tuned and Autoencoder-LSTM generated DT models after completing a single lifelong update step. For instance, when the $i$-th lifelong update step is completed, the MSE values at all future degradation stages are obtained from the corresponding fine-tuned DT and the Autoencoder-LSTM generated DT. The success ratio $sc_i$ at the $i$-th update step is defined as:

$$sc_i = \frac{n^i_{better}}{n^i_{all}} \quad (5)$$

where $n^i_{better}$ is the number of future stages when the Autoencoder-LSTM generated DT has a smaller MSE value than the fine-tuned DT at the $i$-th update step. $n^i_{all}$ is the number of all future degradation stages at the $i$-th update step.

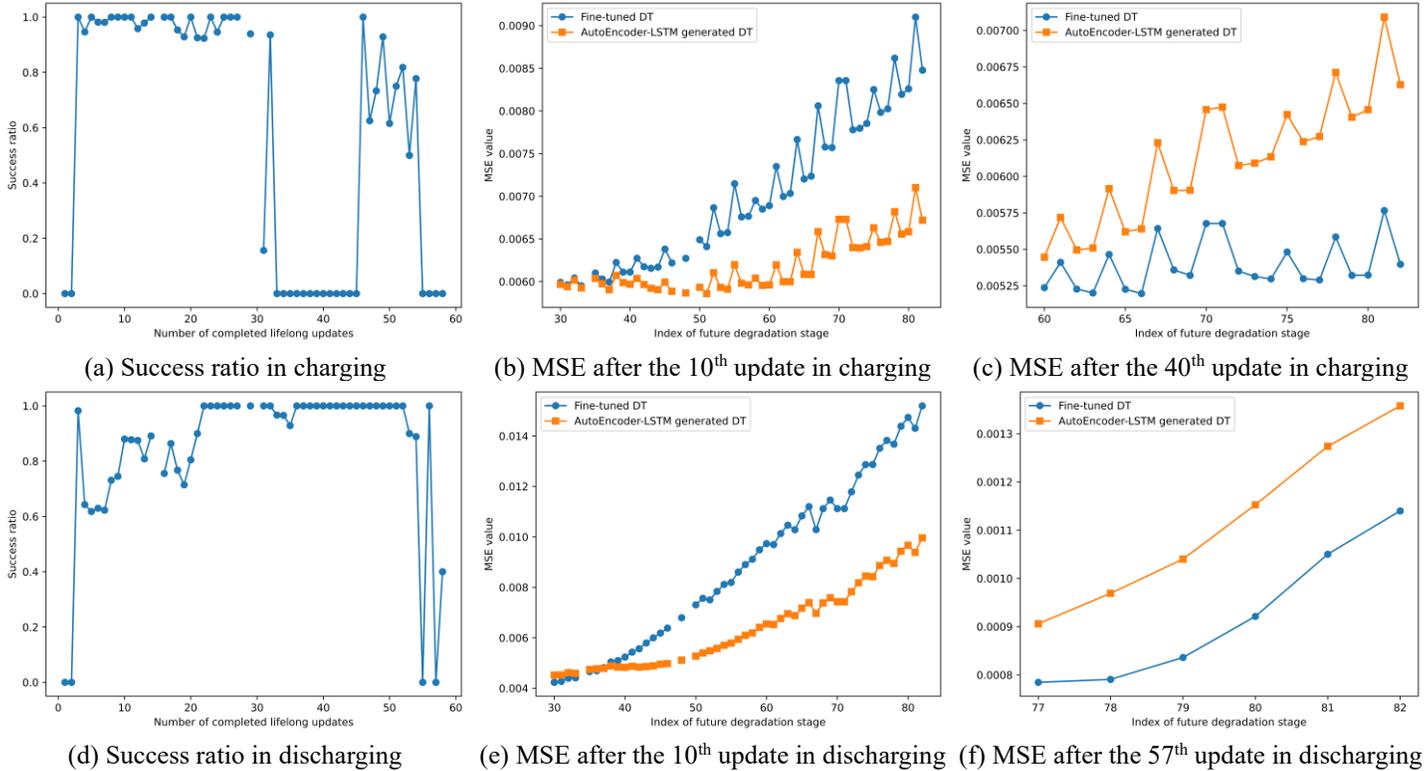

(a) Success ratio in charging  (b) MSE after the 10$^{th}$ update in charging  (c) MSE after the 40$^{th}$ update in charging

(d) Success ratio in discharging  (e) MSE after the 10$^{th}$ update in discharging  (f) MSE after the 57$^{th}$ update in discharging

Figure 13 Success ratio and comparison of MSE curves with the battery dataset.

The success ratios at all lifelong update steps when testing on the charging and discharging datasets are summarized in Figure 13, along with MSE curves at the selected update steps. During the first several lifelong update steps, the success ratio values could be low and even close to zero. This means that the fine-tuned DT has a smaller MSE value than the proposed method at all future degradation stages. This is attributed to the insufficient warm-up phase, which fails to cover enough information on the dynamic behavior of DT configuration. Although the low success ratio could sometimes be observed in the middle of the lifelong update phase as shown in Figure 13 (a) and (d), the difference between the MSE values obtained by fine-tuned and Autoencoder-LSTM generated DT models is quite small. For example, the maximum MSE difference between the two DT models after completing the 40$^{th}$ update is around 0.002 for battery charging in Figure 13 (c), and the max difference after the 57$^{th}$ update is around 0.0001 for battery discharging in Figure 13 (f). It is observed that the success ratio values keep zeros at several successive lifelong update steps for battery charging in Figure 13 (a). The possible reason is that the Autoencoder-LSTM is trained with the constant hyperparameter setting, which would not provide better training performance at some update steps. However, at most lifelong update steps in both tests, the success ratio value is larger than 0.6 as shown in Figure 13 (a) and (d). When the success ratio is close to one, the MSE value obtained by the proposed method is smaller than that obtained by the fine-tuned DT at each future degradation stage, as shown in Figures (b) and (e). Both observations demonstrate that the Autoencoder-LSTM generated DT could provide better predictions for future degradation stages.

Based on the above discussion, the proposed lifelong update method successfully enables the DT model to capture the effects of battery degradation on the battery charging and discharging and outperform the conventional fine-tuning method in terms of prediction performance at future unseen degradation stages in most lifelong update steps.

## 4 NASA Engine Dataset

### 4.1 Dataset description and preparation

The engine dataset covers a group of run-to-failure trajectories for a fleet of aircraft engines under real flight conditions, which are generated by the Commercial Modular Aero-Propulsion System Simulation (CMAPSS) model developed at NASA [61]. During the data generation process, the flight conditions (i.e., altitude $alt$, flight mach number $Mach$, throttle-resolver angle $TRA$, and total temperature at fan inlet $T2$) recorded from a commercial jet are input to the CMAPSS model with the degradation of the engine components. Fourteen measurable time-series physical properties observed by sensors are simulated by the CMAPSS model, which outputs the estimated unobserved health parameters simultaneously. This process repeats as the number of degraded units increases until the health index of the engine reaches the end of its life.

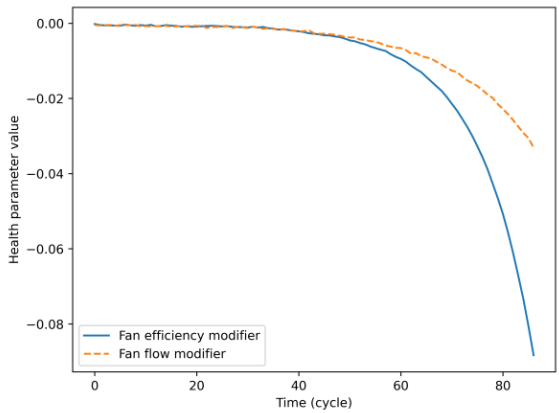
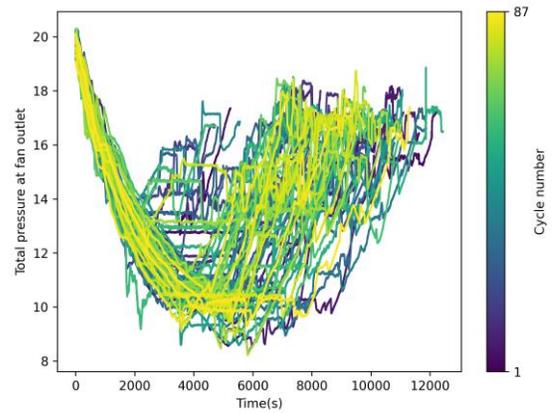

(a) System health indicator          (b) Total pressure $p_1$ at the fan outlet

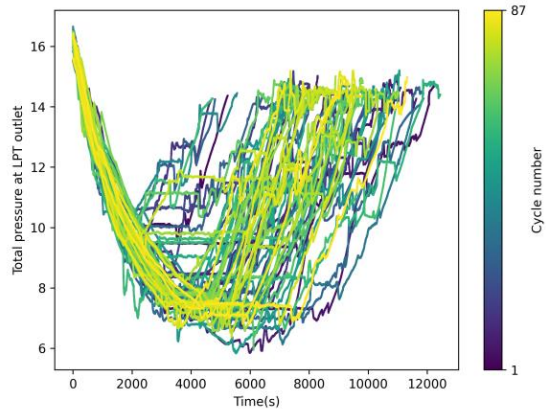

(c) Total pressure $p_2$ at the low pressure turbine (LPT) outlet

Figure 14 Engine degradation curves of unit 1 in dataset DS04.

There are eight datasets with different degradation units in the published repositories, where the health parameter degrades with the flight cycle number in each unit. This study is only tested on the data with the unit number 1 in the file "DS04.h5", where 87 flight cycles are observed in the selected data. Based on the selected dataset, Figure 14 summarizes the degradation of two health parameters over 87 flight cycles and two selected time-series physical properties affected by degradation at each flight cycle. According to Figure 14 (a), both the fan efficiency modifier and the fan flow modifier gradually degrade with the flight cycle. Meanwhile, the time-series total pressures $p_1$ and $p_2$ vary among different flight cycles as shown in Figure 14 (b) and (c), owing to different degradation states at each flight cycle.

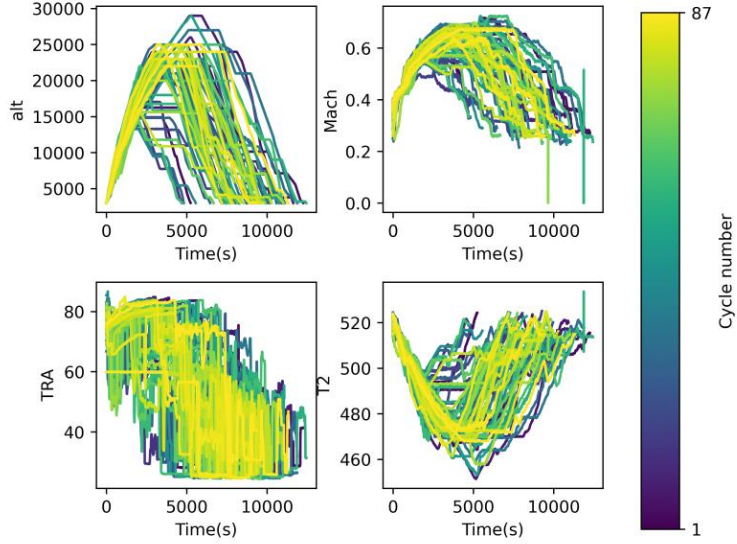

Figure 15 Flight conditions during each flight cycle with the NASA engine dataset.

In this test, the initial degradation stage is defined as the first flight cycle, and the future degradation stages are then defined chronologically, e.g., the stages $T_0, T_1, T_2$ are the flight cycles 1, 2, and 3 respectively. To evaluate the proposed lifelong update method, two DT models are constructed to predict the total pressures $p_1$ and $p_2$ respectively. The input variables of both DT models consist of four flight conditions (i.e., $alt$, $Mach$, $TRA$, $T2$) and the time $t$ during one flight cycle. The four flight conditions at each flight cycle are summarized in Figure 15, where various magnitudes of condition values are observed. The output is the corresponding pressure $p$ at the time $t$ under the given flight condition, as shown in Figure 14 (b) and (c). In this dataset, the number of data points on the raw pressure curve varies from 4700 to 12500 during 87 flight cycles. To reduce the nonlinearity in the raw dataset, both input and output data are smoothed by the moving average filter with a window size of 50. The down sampling with the same window size of 50 is then performed to reduce the number of data points. Finally, all data of each input or output variable at all flight cycles are scaled simultaneously by the Minmax scaler to the range [0,1]. The scaled datasets are used for DT modeling, fine-tuning, and training dynamic models in the proposed method.

### 4.2 Test setting

Considering that the nonlinearity in the engine datasets is higher than that in the battery dataset, a deeper FNN structure is used for DT models. Apart from the input and output layers, the applied structure has seven hidden layers with the activation function Tanh(·), whose dimensions are detailed in Figure 16. The training process of models $DT_0$ at the initial degradation stage $T_0$ for both selected pressures is the same as that in Section 3.2. The training, validation, and testing processes of $DT_0$ are depicted in Figure 17, where they all converge to almost zero after certain epochs. Therefore, the FNN structure selected as the DT model would be sufficient to capture the nonlinear behaviors of total pressures $p_1$ and $p_2$.

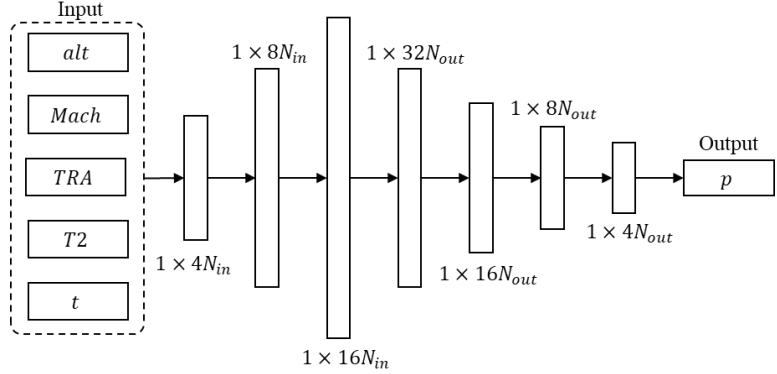

Figure 16 FNN structure for the NASA engine dataset. The input contains the altitude $alt$, flight mach number $Mach$, throttle-resolver angle $TRA$, the total temperature at fan inlet $T2$, and time $t$. The output is the pressure $p$. The input and output dimensions are $N_{in} = 5$ and $N_{out} = 1$ respectively.

As the applied FNN structure is deeper in this test, the entropy values of $\boldsymbol{\theta}_0$ in two initial DT models for both pressures are around 12.9 and the dimensions of latent space in Autoencoders are determined as 26 based on the predefined hyperparameters. During the fine-tuning process of the DT model at each degradation stage, and the training process of the autoencoder and the LSTM, the hyperparameter settings are identical to Table 4. The only difference is that the number $m$ of stages in the warm-up phase and the window size $w$ for input sequences of LSTM are updated as $m = 40$ and $w = 10$ for the engine dataset.

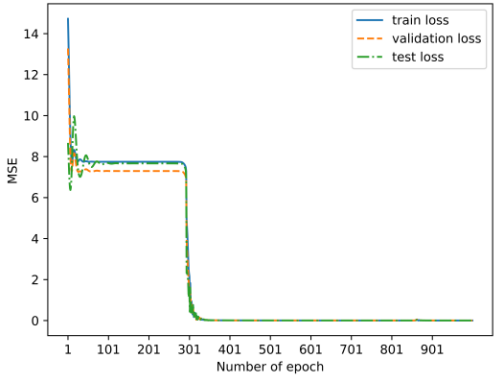
(a) Total pressure $p_1$ at the fan outlet

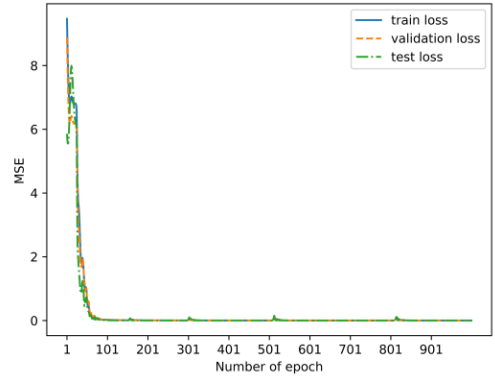
(b) Total pressure $p_2$ at the LPT outlet

Figure 17 Training process of $DT_0$ for the engine dataset.

### 4.3 Results

When assessing the proposed method in the engine dataset, the obtained results are analyzed with the same setting for the battery dataset in Section 3.

After completing each lifelong update step, the MSE values at future degradation stages obtained by the tested two DT models are summarized in Figure 18. From Figure 18 (a) and (c), the MSE values obtained by the fine-tuned DT fail to converge at each lifelong update for both total pressures $p_1$ and $p_2$. Although the MSE values have a large variance in $p_1$ and $p_2$ at the first 10 lifelong updates as shown in Figure 18 (b), the performance of Autoencoder-LSTM generated DT converges during future lifelong updates. Moreover, the MSE values obtained by the proposed method converge at all update steps for total pressure $p_2$, shown in Figure 18 (d). To clarify the performance difference at each degradation stage, the boxplots of MSE values obtained by the two compared DT models are shown in Figure 19, with the same setting as for the battery dataset test. The comparison results show that the Autoencoder-LSTM generated DT has smaller median MSE values and smaller box sizes at each degradation stage. This observation indicates that the proposed lifelong update method outperforms the conventional fine-tuning method in terms of both prediction accuracy and robustness for future unseen stages.

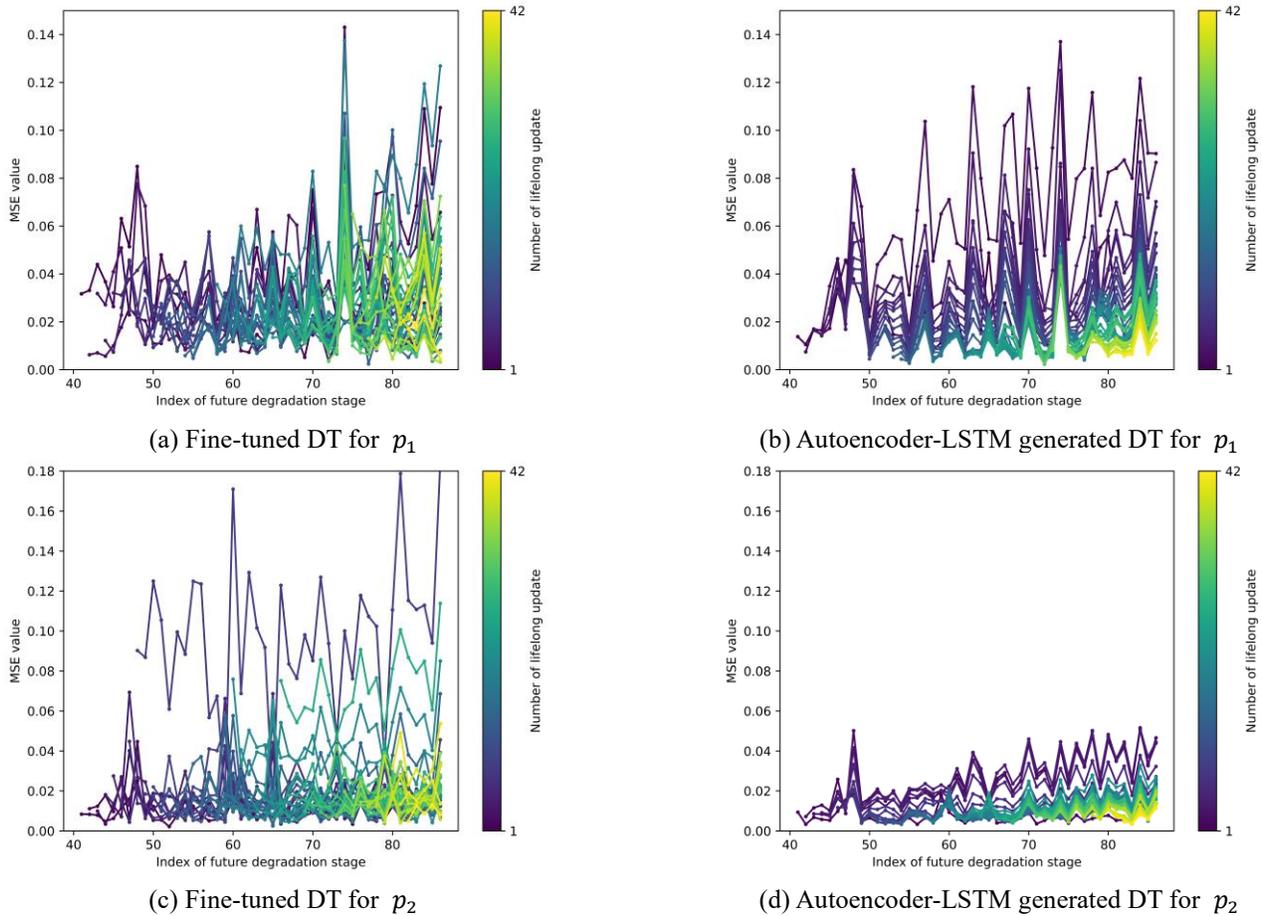

Figure 18 MSE values obtained by fine-tuned and Autoencoder-LSTM generated DT models with the engine dataset. $p_1$ and $p_2$ refer to the total pressure at the fan outlet and the total pressure at the LPT outlet, respectively.

The success ratio calculated at each lifelong update during the test is shown in Figure 20. At most lifelong updates, the value is larger than 0.5 for $p_1$ and $p_2$, as shown in Figure 20 (a) and (d). Meanwhile, in general the success ratio increases with the number of completed lifelong update steps. More specifically, although the ratio value 0 is observed at the second update step for both pressures in Figure 20 (b) and (e), the value could increase to 1 at the 20$^{th}$ update, when the Autoencoder-LSTM generated DT outperforms the fine-tuned DT at each future degradation stage as shown in Figure 20 (c) and (f). Therefore, with the data from more degradation stages, the proposed lifelong update method would learn more information about the dynamic behavior of model configurations during the lifecycle.

Based on those discussions, the proposed lifelong update method could identify the effects of engine degradation on engine responses via learning the trend of model configurations during the lifecycle. Moreover, compared with the conventional fine-tuning framework, the proposed lifelong update method could provide better prediction performance for future unseen degradation stages in most cases.

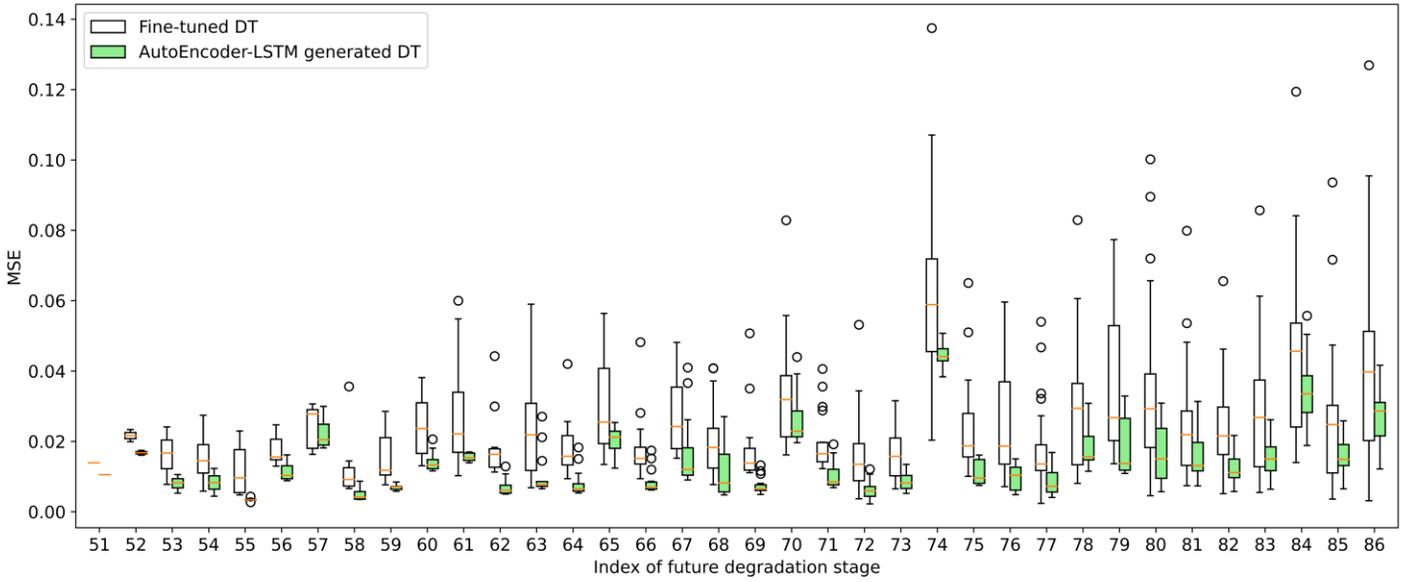

(a) MSE of the total pressure $p_1$ at the fan outlet

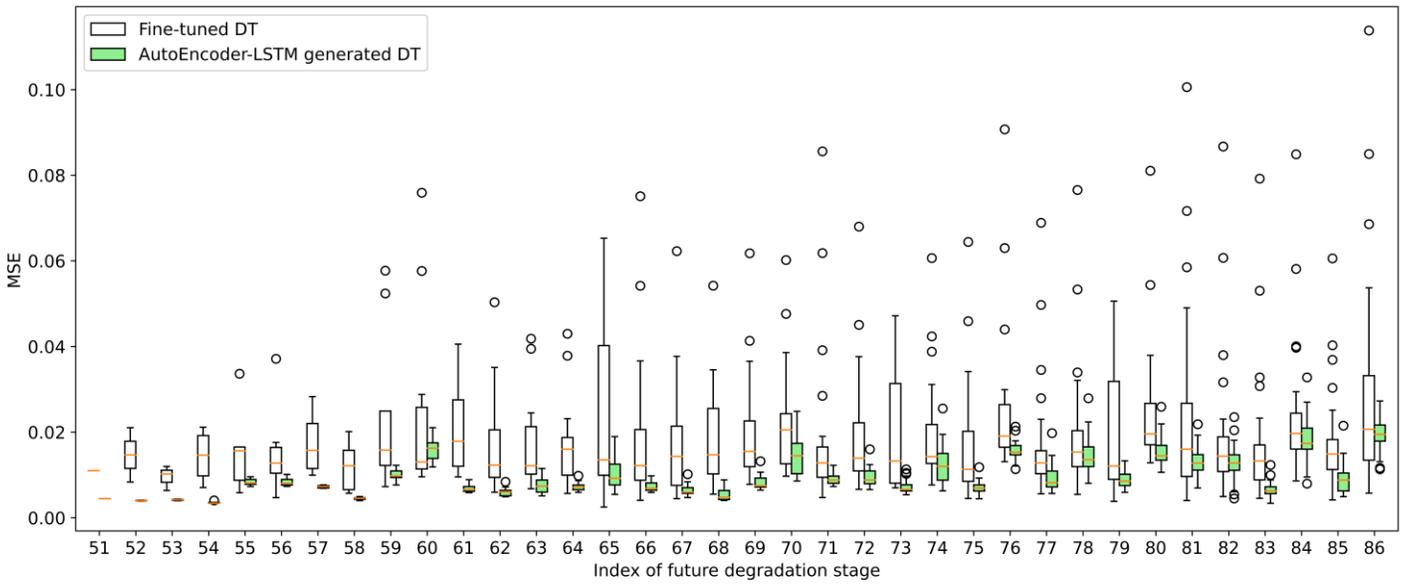

(b) MSE of the total pressure $p_2$ at the LPT outlet

Figure 19 Boxplot of MSE values at each stage during the lifelong update phase with the engine dataset.

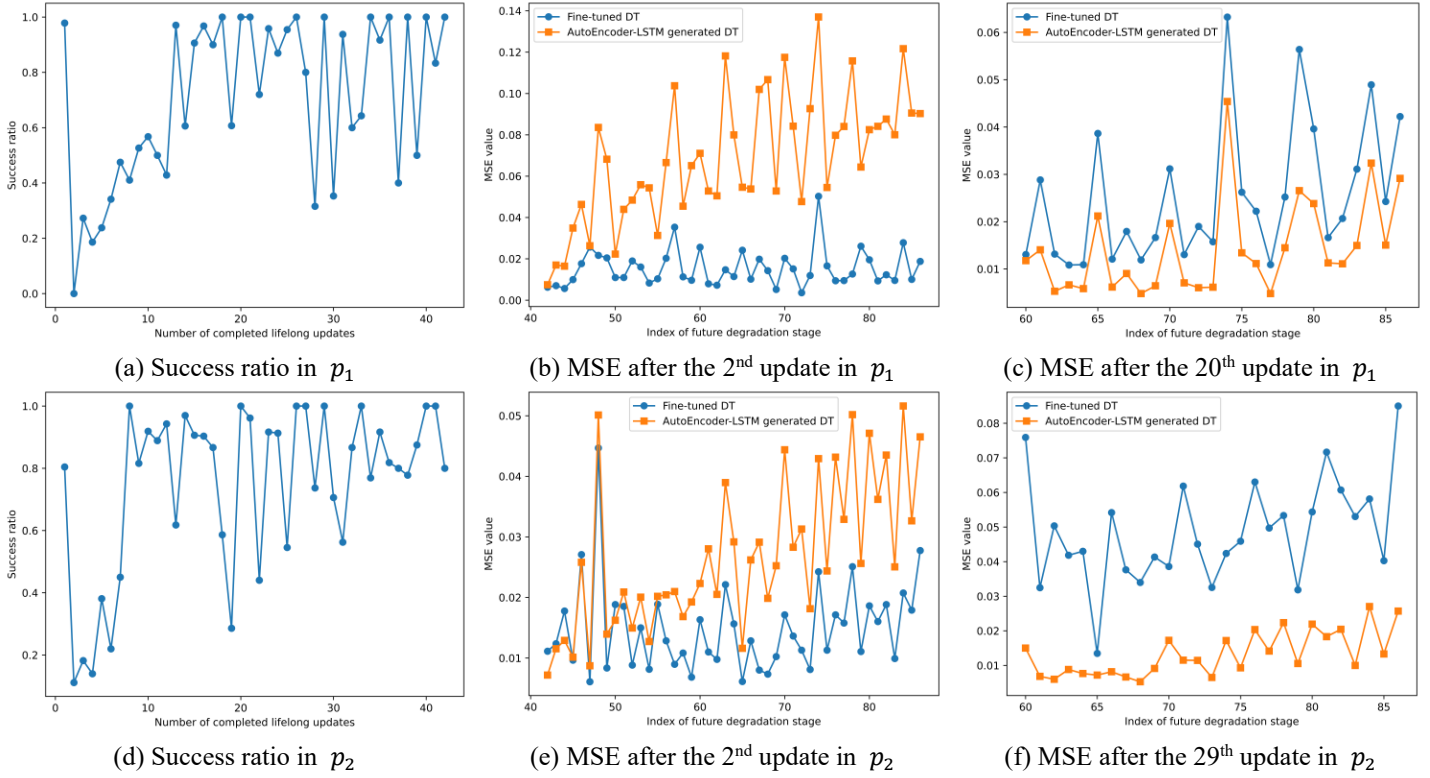

Figure 20 Success ratio and comparison of MSE curves with the engine dataset.

## 5  Discussion and Future Work

### 5.1  Transformer

In both tested real-world problems, only the LSTM model has been applied to capture the changing behavior of model configuration features learned by the proposed Autoencoder. Although the LSTM architecture works well in both studies, there are other model structures in the time-series forecasting field. This section explores the potential of other state-of-the-art model structures for the proposed lifelong update task.

Transformer has been another powerful architecture for time-series forecasting [62, 63] though initially proposed for natural language processing in 2017 [64]. Different from LSTM relying on the recurrence structure and the gating mechanism to process the data sequentially, Transformer adopts the attention mechanism to process all inputted tokens (i.e., data in the time-series sequence) simultaneously, where the importance of different inputs relative to each other is weighted in parallel. The most relevant parts of the input sequences are then identified, ensuring that the Transformer can capture both short-term fluctuations and long-term trends in complex datasets. Due to those merits, Transformer has been adopted for multiple engineering tasks, such as fuel cell degradation prediction [36], energy consumption estimation [65], and weather forecasting [66]. More details of transformer applications and structure variants can be found in literature reviews [62, 63].

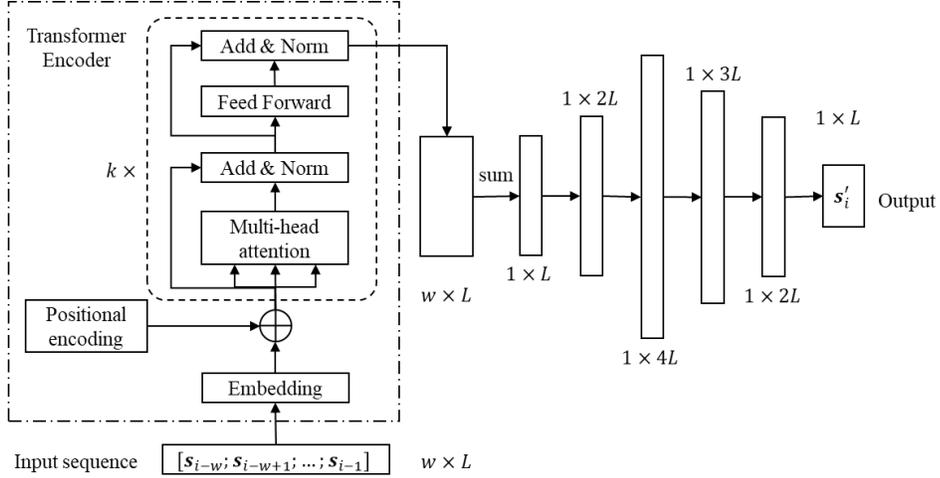

Figure 21 Transformer structure to predict the configuration feature.

The structure of the implemented Transformer in the study is shown in Figure 21. Compared with the LSTM model in Section 2.3.2, the only difference is replacing the 2-layer LSTM block as the Transformer encoder block to capture the encoded features from the input sequence $[s_{i-w}; \ldots; s_{i-1}]$ obtained during the stage window $[i-w, i-1]$. The applied encoder block includes $k$ transformer encoder layers, where each layer has a multi-head attention part and a feedforward model. The encoder block is implemented by the Pytorch library in the work. All $k=6$ encoder layers share the same hyperparameter setting, such as $2L$ as the dimension of the feed-forward network model, 1 as the number of attention heads, and 0 as the dropout rate. The output dimension of the encoder block is the same as the input dimension. A sum operator is then applied to convert the encoder output matrix to the encoded feature vector, whose size is $1 \times L$. Finally, several hidden layers with the activation function $\text{Tanh}(\cdot)$ are assigned to obtain the final output as the predicted latent feature vector $s'_i$ at the next $i$-th stage.

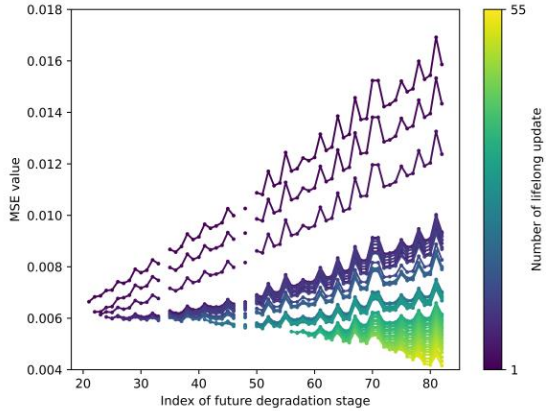

(a) Battery charging

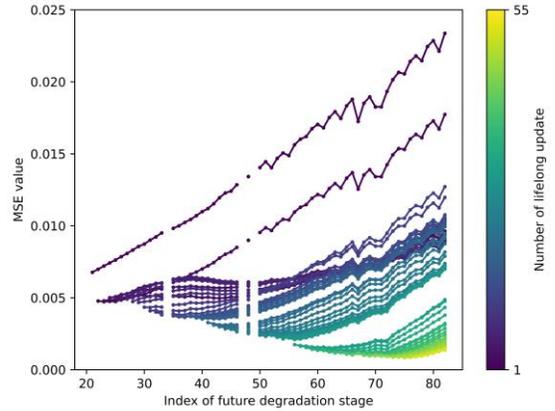

(b) Battery discharging

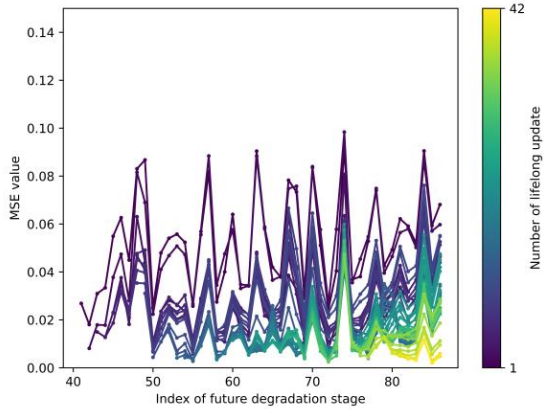
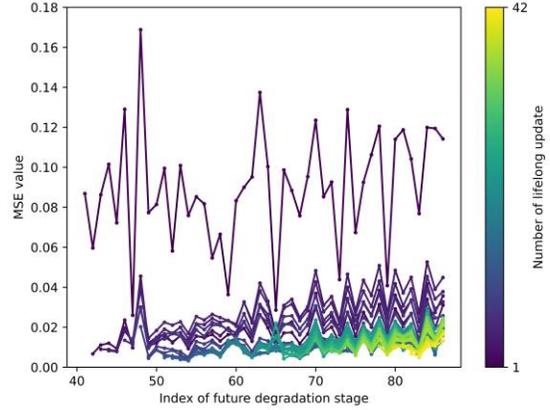

(b) Engine pressure $p_1$               (b) Engine pressure $p_2$

Figure 22 MSE values obtained by Autoencoder-Transformer with the engine dataset.

The implemented Transformer model is integrated into the proposed Autoencoder, which is tested with both the battery degradation dataset and engine dataset. The tested method is denoted as Autoencoder-Transformer in the following discussion for clarification. The training and testing setting for Autoencoder-Transformer is the same as for Autoencoder-LTM in Section 3 and Section 4. After completing one lifelong update with Autoencoder-Transformer, MSE values at all future degradation stages are summarized in Figure 22. Meanwhile, the success ratio at each lifelong update obtained by Autoencoder-Transformer is depicted in Figure 23. Compared with results from Autoencoder-LSTM in Sections 3 and 4, Autoencoder-Transformer provides similar performance. More specifically, both Autoencoder-Transformer and Autoencoder-LSTM may not show better MSE values at the first several lifelong updates than the fine-tuned DT. However, with data from more degradation stages and more updates, they both provide MSE value curves that are lower and more robust than the conventional fine-tuning method at future degradation stages. Although both Autoencoder-Transformer and Autoencoder-LSTM fail to provide better performance than the conventional fine-tuning method with the battery charging task during the latter part of the lifelong update, the success ratio values are larger than 0.5 at most updates and increase gradually with the number of completed update in another three tested problems.

The above discussions demonstrate that the proposed lifelong update method is applicable to be integrated with different time-series forecasting models, which would provide similar performance and outperform the conventional fine-tuning method in terms of performance and robustness for future unseen degradation stages.

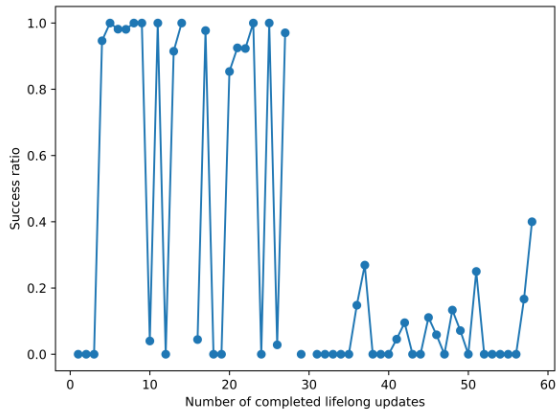
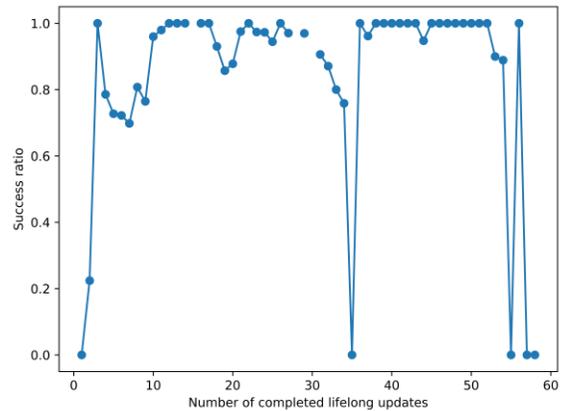

(a) Battery charging               (b) Battery discharging

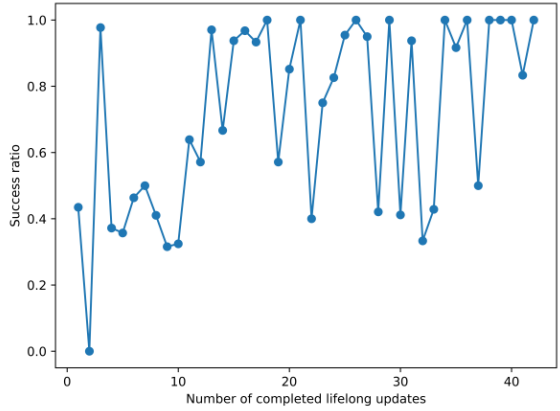
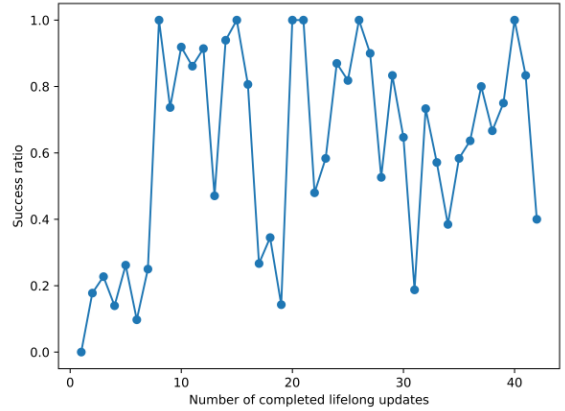

(b) Engine pressure $p_1$                                            (b) Engine pressure $p_2$

Figure 23 Success ratio obtained by Autoencoder-Transformer with the engine dataset.

## 5.2 Effect of Autoencoder

In this work, parameters of all hidden layers in the FNN model are fed to the proposed Autoencoder together to capture potential dependencies among hidden layers, as discussed in Section 2.3.1. To further study the effect of the applied option, this section considers each hidden layer independently. That means that for an FNN model with $N$ hidden layers, $N$ AutoEncoder models are trained to reconstruct the parameters of each layer respectively. Then $N$ LSTM models are adopted to capture the changing behavior of learned parameter features during the lifecycle for $N$ hiden layers respectively. During the test, the Autoencoder structure of the $i$-th layer is shown as Figure 24, where the output of each hidden layer is activated by Tanh(·). The LSTM model adopted in the test is the same as that in Section 2.3.2. The training and testing setting are identical to the results in Section 3 and Section 4. To clarify the following discussion, the proposed model in Section 2 is denoted as Autoencoder-LSTM *per model*, and the tested method in this section is denoted as Autoencoder-LSTM *per layer*.

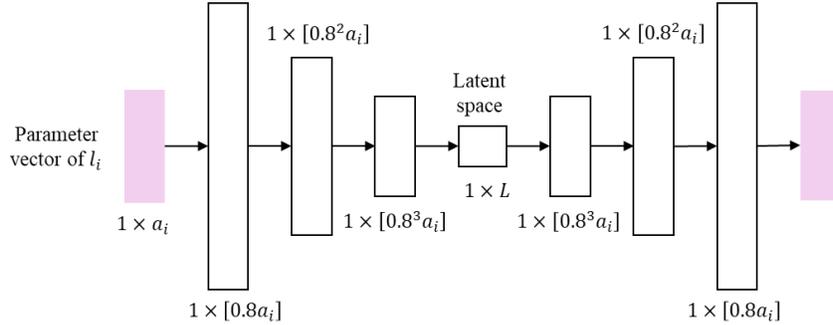

Figure 24 Autoencoder structure for single hidden layer. $a_i$ is the number of parameters in the hidden layer $l_i$. [·] refer to the nearest integer.

The results obtained by Autoencoder-LSTM per layer for the battery charging dataset and the engine pressure $p_1$ are summarized in Figure 25. Compared with results of Autoencoder-LSTM per model in Figure 11 and Figure 18, the MSE values from Autoencoder-LSTM per layer fail to converge to certain values over the degradation stages in battery charging (Figure 25 (a)), where the MSE value always increases with the number of future degradation stage and is larger than that obtained by Autoencoder-LSTM per model. Meanwhile, even after some lifelong updates, the tested model still provides a performance with larger variance in engine pressure dataset, where the MSE value is larger than 0.02 at most degradation stages in Figure 25 (b). In contrast, the MSE values obtained by Autoencoder-LSTM per model are below 0.02 after some lifelong updates, as shown in Figure 18

(b).

Based on the above comparison, feeding parameters of all hidden layers to Autoencoder simultaneously would benefit learning the latent features that consider dependencies among hidden layers. Such operation outperforms the alternative considering each hidden layer independent in terms of accuracy and robustness at the final reconstructed DT models.

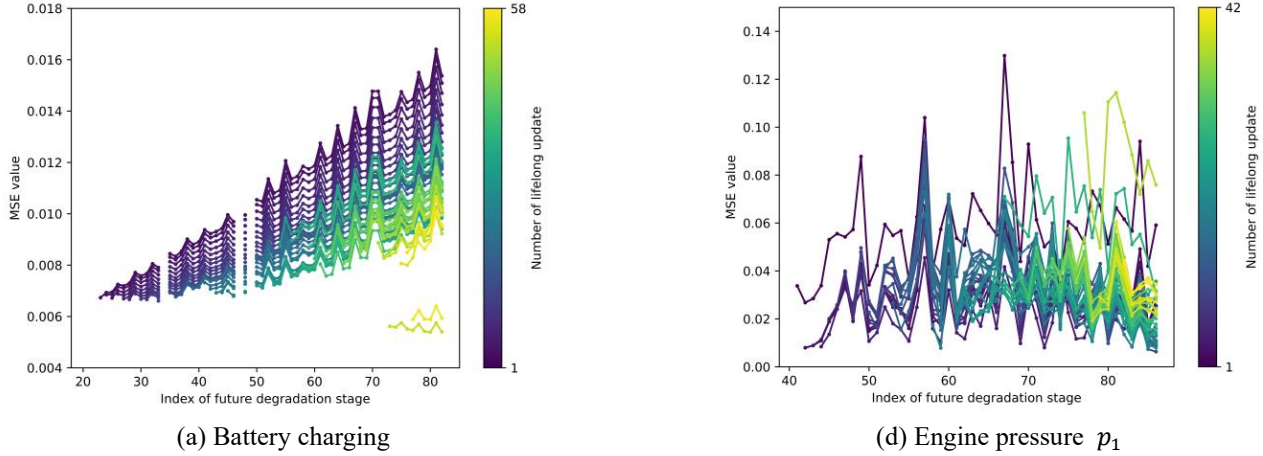

(a) Battery charging　　　　　　　　　　　　　　(d) Engine pressure $p_1$

Figure 25 Results of Autoencoder-LSTM per layer in battery charging and engine pressure $p_1$

## 5.3 Future Work

Although the benefits of the proposed lifelong update method have been demonstrated by real-world battery and engine datasets, the degradation behavior exist in many practical systems, such as bearing [67], gear [22], and infrastructure [23]. The generalizability and applicability of the proposed lifelong update method needs more tests on lifecycle datasets from various engineering fields.

In the work, the DT model structure is assumed to be fixed at each degradation stage. However, in real-world problems, the nonlinearity and the nonstationary of data would change with the degradation stages in the system lifecycle. In such cases, the optimal model structure also evolves with the degradation stage. A similar case is observed in the electro-optical system, where a nonparametric Bayesian graph neural network is constructed to denote the dynamic degradation process of health stage and the propagation of epistemic uncertainty [14]. When the data at new time steps are observed, the Dirichlet process mixture model and the Gaussian particle filter are adopted to update the optimal network structure and obtain the new model parameters respectively. However, the work [14] only updates the model step by step, still failing to find the optimal model structure at future time steps in advance. Therefore, one future work is to find a method to represent the dynamic behavior of optimal model structures and corresponding parameters simultaneously over the degradation stages in the lifecycle, which enables predicting the optimal configuration (i.e., structure, and parameters) at unseen future stages.

Moreover, constant training hyperparameters are utilized at each degradation stage. This would not be the best operation, as the number of data and the data structure change gradually. This may contribute to the observation that both Autoencoder-LSTM and Autoencoder-Transformer show worse performance at some degradation stages during the latter half lifelong update phase, as shown in Figure 13 (a) and Figure 23 (a). Therefore, how to choose the optimal hyperparameter automatically for learning at each stage would be another future work to make the DT provide stable and better performance during the lifelong update.

## 6　Summary

For systems degrading during their lifecycle, this paper proposes a lifelong update method for corresponding digital twin models. Different from conventional online digital twin update focusing on performance only at one degradation stage, the proposed method aims to predict the system responses affected by system degradation at all unseen future degradation stages. To fulfill this purpose, the system degradation during the lifecycle is represented as the dynamic behavior of the digital twin configuration at continuous degradation stages. During the lifelong update process, the digital twin model, i.e., a feedforward neural network with the fixed structure in this work, is fine-tuned as a conventional method at each degradation stage. The optimized network parameters at all known stages are stored to train an Autoencoder model to learn the latent features that involve the dependencies among all hidden layers. Based on the obtained latent features at known degradation stages, a long short-term memory model is adopted to identify the dynamic behavior of latent features over degradation stages. Once both models are trained, the latent features at one future unseen degradation stage could be estimated by the long short-term memory model, and the network parameters at the same stage are reconstructed by the decoder part in Autoencoder from the estimated latent features. The digital twin model is finally determined to forecast the system responses affected by potential degradation at the same stage. The benefits of the proposed method have been demonstrated by test results in two real-world engineering datasets, i.e., battery degradation data, and flight engine dataset. Compared with the conventional fine-tuning method, the proposed method can provide predictions with better accuracy and robustness at future unseen degradation stages.

Meanwhile, Transformer, another powerful tool in time-series forecasting task, is compared with long short-term memory model in the proposed method. Both methods provide similar performance, indicating that the proposed method is adjustable with different time-series forecasting models. Moreover, processing the parameters of all hidden layers of the DT model simultaneously facilitates the Autoencoder finding the latent feature involving relationships among all hidden layers. Finally, discussions of future works are presented in terms of engineering datasets for generalizability tests, updating the model structure and parameters simultaneously, and automatic strategy to obtain the optimal training hyperparameter setting at each degradation stage.

With the proposed lifelong update method, a DT model is able to identify how the system responses are affected by system degradation during the lifecycle, even though the system is far from its end of life. The proposed method provides an economical alternative to expensive aging experiments and enables a system to deploy to work directly without any prior aging tests. This would bring some economic benefits to system manufacturers. Meanwhile, it can predict the responses of degradation-affected systems at unknown future stages, which is beneficial to conducting health management and monitoring.

**Acknowledgements**

Funding from Natural Sciences and Engineering Research Council (NSERC) of Canada under the project RGPIN-2019-06601, as well Graduate Dean's Entrance Scholarship (GDES) from Simon Fraser University, is appreciated.